\documentclass[12pt]{article}

\usepackage[dvips]{epsfig}

\usepackage{amsmath,amssymb,color,graphics}
\usepackage{bm}% bold math
\usepackage[%breaklinks=true,
colorlinks=true,backref,pagebackref]{hyperref} 

\usepackage[stable]{footmisc}

\definecolor{dark-green}{rgb}{0,0.7,0}
\definecolor{dark-blue}{rgb}{0,0.2,0.5}
\definecolor{med-blue}{rgb}{0,0.7,1}
\definecolor{mblue}{rgb}{0,0.2,1}
\definecolor{cnc}{rgb}{0.8,0,0}
\definecolor{light-red}{rgb}{1,0.8,0.8}
\definecolor{dark-yellow}{rgb}{1,0.8,0}
\definecolor{light-blue}{rgb}{0.8,0.9,1}
\definecolor{verylight-blue}{rgb}{0.93,0.95,1}
\definecolor{light-yellow}{rgb}{1,0.9,0.8}
\definecolor{grey}{gray}{0.88}

\def\a{\alpha}

\def\d{\delta}

\def\vp{\varphi}

\usepackage{amsfonts}

\begin{document}

\title{On Kottler's path: origin and evolution of the premetric
  program in gravity and in electrodynamics\footnote{Based on an
    invited talk given at the Annual Meeting of the German Physical
    Society (DPG) in Berlin on 20 March 2015, ``Working Group on
    Philosophy of Physics (AGPhil)''. We sincerely thank Dennis
    Lehmkuhl (Wuppertal, now Pasadena) and his co-organizers Meinard
    Kuhlmann (Mainz) and Wolfgang Pietsch (Munich)
    \cite{Kuhlmann:2012} for the invitation. In the present version,
    we appreciably enriched our text by adding more formalism. We
    apologize to the philosophers for any inconvenience. A short
    version of this paper will be submitted to the International
    Journal of Modern Physics D.}}

\author{Friedrich W. Hehl$^{1}$, Yakov Itin$^{2}$, Yuri
  N.~Obukhov$^{3}$\\ \\
  $^{1}$Inst.\ for Theor.\ Physics, Univ. of Cologne, Germany\\ and
  Dept.\ Physics \& Astron., Univ.\ of Missouri, \\
  Columbia, MO, USA; hehl@thp.uni-koeln.de\\
  $^{2}$Inst.\ of Math.\ Hebrew Univ.\ of Jerusalem and\\
  Jerusalem College of Technology, Israel;\\itin@math.huji.ac.il\\
  $^{3}$Nuclear Safety Inst., Russian Academy of Sciences,\\ Moscow,
  Russia; obukhov@ibrae.ac.ru}

\date{}
\maketitle
\vspace{-20pt}
\begin{abstract}In 1922, Kottler put forward the program to remove the
  gravitational potential, the metric of spacetime, from the {\it
    fundamental} equations in physics as far as possible. He
  successfully applied this idea to Newton's gravitostatics and to
  Maxwell's electrodynamics, where Kottler recast the field equations in
  premetric form and specified a metric-dependent constitutive law.
  We will discuss the basics of the premetric approach and some of its
  beautiful consequences, like the division of universal constants
  into two classes. We show that classical electrodynamics can be
  developed without a metric quite straightforwardly: the Maxwell
  equations, together with a local and linear response law for
  electromagnetic media, admit a consistent premetric formulation.
  Kottler's program succeeds here without provisos. In Kottler's
  approach to gravity, making the theory relativistic, two premetric
  quasi-Maxwellian field equations arise, but their field variables,
  if interpreted in terms of general relativity, do depend on the
  metric. However, one can hope to bring the Kottler idea to work by
  using the teleparallelism equivalent of general relativity, where
  the gravitational potential, the coframe, can be chosen in a
  premetric way.

  \hfill {\it file Kottler2015IJMPD26.tex, 21 July 2016}
\end{abstract}
%\newpage
\vspace{2truecm}

\newpage
\tableofcontents\vspace{2.5truecm}

%\newpage
\hfill
\begin{minipage}[c]{10cm} {\it Since the notion of metric is a
    complicated one, which requires measurements with clocks and
    scales, generally with} rigid {\em bodies, which themselves are
    systems of great complexity, {it seems undesirable
      to take metric as fundamental,} particularly for phenomena which
    are simpler and actually independent of it. 

\hfill Edmund Whittaker
    (1953)}%\medskip

%\hfill [emphasis by us]
\end{minipage}

%%%%%%%%%%%%%%%%%%%%%%%%%%%%%%%%%%%%%%%%%%%%%%%%%%%%%%%%%%%%%%%%%%%%%
\section{Introduction}\label{Intro}
%%%%%%%%%%%%%%%%%%%%%%%%%%%%%%%%%%%%%%%%%%%%%%%%%%%%%%%%%%%%%%%%%%%%%

The physical reality of the metric can, as noticed by van Dantzig
\cite{vanDantzig:1955}, hardly be ``...denied by anyone who has ever
been pricked by a needle, i.e. who has {\it felt\/} its rigidity and
the smallness of its curvature.''  Yes, we felt this, too,
repeatedly. That is, the reality of the metric is beyond doubt. Still,
as Whittaker observed, there are phenomena in physics that are
independent of the metric or for which the metric turns out to be
irrelevant. If you count the number of electrons within a prescribed
domain, for example, this number does not refer to a length
measurement---and a domain, a container, can be defined without the
necessity to have a length measure available.

Historically, the {\em premetric program} in physics was first clearly
spelled out by Friedrich Kottler in 1922 in two papers on ``Newton's
law and metric'' and on ``Maxwell's equations and metric''
\cite{Kottler:1922a,Kottler:1922b}. A short history of the subject is
provided in Whittaker's book \cite[Vol.2, pp.\ 192--196]{Whittaker:1953}
to which we refer for more details. We will discuss Kottler's two
papers in a modern formalism and will show that the premetric nature
of Newton's (non-relativistic) gravitational attraction law is
phenomenologically not as well grounded as that of the Maxwell
equations. However, in relativistic theories of gravity, the premetric
program leads to a better understanding of the interdependence between
topological, affine, and metric concepts, that is, between those
concepts related to the surgery of a manifold, to parallel transfer,
and to spatio-temporal distances, respectively.

In van Dantzig's nomenclature
\cite{vanDantzig:1955,vanDantzig:1934,vanDantzig:1934/6,vanDantzig:1937},
the premetric Maxwell equations are called ``fundamental equations''
whereas the set of metric-dependent ``linking equations'' relates the
excitations to the field strengths, and the potentials to the currents
(typically by Ohm's law). The linking equations depend on the
properties of the vacuum or of the matter involved. They can be
adapted to the structure of the ``material'' considered, whereas the
fundamental equations are, on the classical level, unalterable.

This qualitative difference between premetric and metric quantities is
also reflected in the universal constants of physics; for an overview
of the fundamental constants, see Flowers \& Petley \cite{Flowers} and
Mohr et al.\ \cite{CODATA:2012,Mohr:2015ccw}. There are universal scalars,
like the Planck constant $\hbar$, the elementary charge $e$, the
magnetic flux quantum $\Phi_0$, which are invariant under all
coordinate transformations (that is, under the diffeomorphism
group). They all can be counted. Today single electrons, for example,
can be manipulated by nanotechnological means. In contrast, the speed
of light is only a P(oincar\'e)-scalar, cf.\ Fleischmann
\cite{Fleischmann:1971}. It is merely invariant under the rigid
Poincar\'e transformation of the flat Minkowski space of special
relativity. Since a speed requires for its measurement spatial and
temporal distances, the speed of light $c$ is metric-dependent and is,
in this sense, less fundamental than $\hbar,e,\Phi_0$. We all know
this since the deflection of light near the Sun has been
experimentally confirmed. In other words, $c$ is only ``universal''
provided gravity can be neglected. Since in our terrestrial
laboratories gravity is relatively weak, it can be neglected if $c$ is
measured.

In Secs.\,2 to 5, we display, in a modern disguise, Kottler's
electrodynamic results of 1922. In Sec.\,2, a premetric 3-vector
calculus is set up. In Sec.\,3, we start {}from a 4-dimensional premetric
version of the Maxwell equations in the calculus of exterior
differential forms and put the Maxwell equations into the formalism of
Sec.\,2; for textbook representations, one could compare with Schouten
\cite{Schouten:1989}, Bamberg \& Sternberg \cite{Bamberg:1990}, and
Scheck \cite{Scheck}.

In Sec.\,4---this section can be skipped at a first reading---the jump
condition at the boundaries between different materials are directly
derived {}from the premetric Maxwell equations. In Sec.\,5, the
Maxwell-Lorentz constitutive law for vacuum electrodynamics is
presented. Here, for the first time, the metric of spacetime enters
electrodynamics. This concludes our review of Kottler's results
\cite{Kottler:1922b} of 1922 on electrodynamics.

Sec.\,5 immediately invites to a reflection about the physical
dimensions and units in electrodynamics and on the fundamental
universal constants involved. This is done in Sec.\,6, in particular the
status of the electric constant $\varepsilon_0$ and the magnetic
constant $\mu_0$ are elucidated and the Josephson and the von Klitzing
constants put into a proper perspective.

Next we turn to Kottler's version of Newton's gravity. In Sec.\,7 the
corresponding results of Kottler are presented in 3-vector calculus
and in exterior differential forms. We are not aware of other
literature on these thought-provoking considerations of
Kottler.

Subsequently, in Sec.\,8, we can extend these gravitoelectric results
of Kottler {by introducing gravitomagnetism. We} will
follow ideas that were pronounced already by Heaviside
\cite{Heaviside:1893a}. This will us lead to
{quasi-Maxwellian gravitational field equations, which will
  be interpreted in terms (i) of {\it general relativity} and (ii) of
  the {\it teleparallelism theory} of gravity.  In the former approach
  gravitomagnetism defies a proper premetric framework, in the latter
  approach, however, the gravitational field variable is premetric
  (namely the coframe) and the constitutive assumption seems to be the
  requirement of the coframe to be orthonormal. This appears to be in
  the sense of the Kottler idea.}

In Sec.\,9 we will postulate a premetric local and linear response law
for classical electrodynamics, starting {}from ideas developed in
Sec.\,6.4. This law will thoroughly be studied in Secs.10 and 11. In
particular, we will find out that we can {\it derive} a conformal
metric by insisting on the vanishing of birefringence in the medium
under considerations. In this way, we found a new way to understand
the emergence of the concept of a metric of spacetime. 

In Sec.\,12 we will show that even the signature of the metric can be
understood this way and that it is related to the sign of the
electromagnetic energy and the Lenz rule in the context of Faraday's
induction law. In Sec.\,13 we conclude with some remarks.\medskip

{\sl A short curriculum vitae of Friedrich Kottler (FK)} ($^*$Vienna
1886, $^\dagger$Roche\-ster, New York 1965): FK was born as a son of a
Lutheran attorney-at-law and a Jewish mother. He was a theoretical
physicist, who first worked in relativity theory. He got a Ph.D.\ in
1912 on spacetime lines in Minkowski's world.\footnote{Kottler applied
  the four-dimensional tensor analysis (of Ricci-Curbastro and
  Levi-Civita \cite{Ricci}) to relativity theory even before Einstein
  became aware (via Grossmann) of this tool.} {}From 1914 to 1918, FK
served as an officer in the Austrian army during the first World
War. Already in 1918, he wrote an extended paper on a detailed
analysis of Einstein's gravitational theory \cite{Kottler:1918}. In
1922 he studied the role of the metric tensor in Newton's
gravitational theory and in electrodynamics. During the late 1920s, FK
turned to optics, inter alia, and till 1938 was an academic teacher
and Professor at the University of Vienna.

In 1938, apparently because of political reasons and of his Jewish
mother, he was fired {}from his position by the Nazis (Austria and
Germany were united at this time), and he emigrated to the US. There,
he worked in industry with the Eastman Kodak Company, once a dominant
player in the photographic film sector.\footnote{Around 2000, one of
  us (fwh) met an American physicist (HJZ), who worked in FK's
  laboratory. Apparently, FK was the boss of some major laboratory in
  the Eastman Kodak Company.}  FK became an American citizen in
1945. Around 1955, FK wanted to retire at Kodak and tried to get his
University position back. The University of Vienna made him an
honorary professor of mathematical physics, but FK informed the
University that he could not teach before the Winter term of
1956/57. The teaching of FK in Vienna, however, apparently never
materialized.

A short CV of FK up to 1938 is available \cite{Kottler:1938}, written by FK
himself. It was sent to Einstein and Pauli in order to help FK to find
a position in the US. Havas \cite{Havas:1999}, in a monograph by
Goenner, Renn, Ritter, and Sauer \cite{Goenner:1999}, gave a very
insightful review of the development of the relativity theory in
Austria {}from 1900 to 1938, Therein the role of Kottler is also clearly
visible. An obituary of FK of the Vienna University can be found under
\cite{Kottler:obituary}.

%A more extended version of this paper, including a short cv of
%Friedrich Kottler, can be found on arXiv.org:...."
%

%%%%%%%%%%%%%%%%%%%%%%%%%%%%%%%%%%%%%%%%%%%%%%%%%%%%%%%%%%%%%%%%%%%%%
\section{Premetric 3-vector calculus\footnote{Our more philosophically
    inclined readers may not want to follow the erection of the vector
    calculus and the setting up of the premetric Maxwell equations,
    including the constitutive law, in detail. We collected for them
    the main results of Secs.\,2, 3, and 5 in Table I, Table II, Table
    III and in Figures 1, 2, which they may want to turn to
    directly.}  }\label{poorvect}
%%%%%%%%%%%%%%%%%%%%%%%%%%%%%%%%%%%%%%%%%%%%%%%%%%%%%%%%%%%%%%%%%%%%%

Although one can hardly overestimate the importance of the metric in
physics, it is remarkable how far one can go in the development of
meaningful differen\-tial-geometrical tools for direct applications in
field theory.

We begin with the discussion of a 3-dimensional space $M_3$. We will
{\it not} assume any Riemannian metric to be defined on it. Here we
will demonstrate that, despite a widely spread belief, one can
construct, essentially, a 3-vector calculus without any metric.

Let $x^a$, with $a=1,2,3$, be local coordinates in some chart that we
choose in such a way that the vectors
$\partial_a:={\frac{\partial}{\partial x^a}}$ are linearly independent
at any point of the chart. Thus, they form a coordinate basis of the
tangent space. Consequently, the three 1-forms $dx^a$ comprise a basis
of the cotangent space at any point, being dual to the coordinate
frame $\partial_a$, that is, $\partial_a \rfloor dx^b=\delta^b_a$,
where $\rfloor$ denotes the interior product of a vector with a
form. Since $dx^a$ is a
coframe, the 3-form $\epsilon:= dx^1\wedge dx^2\wedge dx^3$ is defined
in this local chart. One immediately notices, though, that $\epsilon$
is {\it not} invariant under the change of coordinates. Indeed, under
a transformation $x^a\rightarrow x^a(x'{}^b)$, we have
\begin{equation}
\epsilon\longrightarrow\epsilon' = J^{-1}\epsilon,\quad\quad
J:=\det\left({\frac {\partial x^a} {\partial x'{}^b}}\right).\label{eps}
\end{equation}
Thus, the 3-form $\epsilon$ is a {\it density} of the {\it weight}
$-1$.

There are more densities of various (exterior) ranks. Namely, we
define the set of three 2-forms by $\epsilon_a := \partial_a\rfloor
\epsilon$, the set of three 1-forms $\epsilon_{ab} := \partial_b
\rfloor \epsilon_a$, and, finally, an object $\epsilon_{abc}
:= \partial_c \rfloor\epsilon_{ab}$.  {}From (\ref{eps}) it is clear
that these forms are all densities of weight $-1$. At the same time,
$\epsilon_a$ is a covector-valued 2-form, $\epsilon_{ab}$ an
antisymmetric 2-tensor-valued 1-form, and $\epsilon_{abc}$ is a
totally antisymmetric 3-tensor density.

{}From the definition $\epsilon = dx^1\wedge dx^2\wedge dx^3$, we have
$\epsilon_{123} = 1$.  Other components of $\epsilon_{abc}$ are equal
$\pm 1$ for even/odd permutations of the indices 1,2,3, or zero when
any two indices coincide. Moreover, the components of this object
(Levi-Civita symbol) have the {\it same values in all local
  coordinates}. Indeed, under a transformation $x^a\rightarrow
x^a(x'{}^b)$ we find
\begin{equation}
  \epsilon_{abc}\longrightarrow\epsilon'_{abc} = J^{-1}{\frac
    {\partial x^d}    {\partial x'^a}} {\frac {\partial x^e} 
    {\partial x'^b}} {\frac {\partial x^f} {\partial x'^c}}\epsilon_{def}
  = \epsilon_{abc}.\label{epsinv}
\end{equation}
We can uniquely define a similar object $\epsilon^{abc}$ as the
solution of the equation $\epsilon^{abc} \epsilon_{mnk}=
\delta^{[a}_{m} \delta^b_n\delta^{c]}_{k}$. This is a density of
weight $+1$, and the brackets $^{[\;\,]}$ denote antisymmetrization of
the indices involved, see Schouten \cite{Schouten:1989}.

One may wonder of how to use these objects. With their help we can, to
a certain extent, circumvent the absence of metric in $M_3$ and
construct a substitute of the full-fledged vector calculus. We call it
{\it the poor man's vector calculus,} since at one's disposal there
are only the operations of the exterior $\wedge$ and the interior
$\rfloor$ products and of the exterior differential $d$.

To begin with, let us consider an arbitrary 1-form (or covector)
$\omega$.  In local coordinates it is represented by its components
$\omega_a$ according to $\omega = \omega_a dx^a$. It is clear that we
can work equally well with the {\it components} $\omega_a$, keeping in
mind their properties under coordinate transformations.  Let us now
calculate the exterior differential $d\omega$. In local coordinates it
reads:
\begin{eqnarray}\nonumber
%\begin{split}
  d\omega = d\omega_a\wedge dx^a &=& (\partial_2\omega_3
  - \partial_3\omega_2)dx^2\wedge dx^3 +  (\partial_3\omega_1
  - \partial_1  \omega_3)dx^3\wedge dx^1   \\ & & \hspace{-6pt} +
  (\partial_1\omega_2 - \partial_2\omega_1)dx^1\wedge dx^2.\label{rot1}
%\end{split}
\end{eqnarray}
The vector valued 2-form density $\epsilon_a$ forms the basis in the space 
of 2-forms. It has the components
\begin{equation}
\epsilon_1=dx^2\wedge dx^3,\quad \epsilon_2=dx^3\wedge dx^1,\quad
\epsilon_3=dx^1\wedge dx^2.\label{epsA}
\end{equation}
Comparing (\ref{rot1}) with (\ref{epsA}), we can define a differential
operator ${\rm curl}_\epsilon$ which maps covectors into vector
densities of the weight $+1$. In the exterior form language it is
defined implicitly by $\left({\rm
    curl}_{\epsilon}\,\omega\right)^a\epsilon_a:=d\omega$ or
explicitly in components by
\begin{align}\label{rot}
  \left({\rm curl}_{\epsilon}\,\omega\right)^1 &:=
  \partial_2\omega_3 - \partial_3\omega_2,\nonumber\\
  \left({\rm curl}_{\epsilon}\,\omega\right)^2 &:=
  \partial_3\omega_1 - \partial_1\omega_3,\\
  \left({\rm curl}_{\epsilon}\,\omega\right)^3 &:=
  \partial_1\omega_2 - \partial_2\omega_1.\nonumber
\end{align}

One can check directly that $\left({\rm
    curl}_{\epsilon}\,\omega\right)^a$ transforms as
\begin{equation}
  \left({\rm curl}_{\epsilon}\,\omega\right)^a\longrightarrow
  \left({\rm curl}_{\epsilon}\,\omega\right)'{}^a =J\left({\frac {\partial 
        x'{}^a}{\partial x^b}}\right)\left({\rm
      curl}_{\epsilon}\,\omega\right)^b,
\end{equation}
that is, like a vector density of the weight $+1$.

Likewise, given an arbitrary 2-form $\varphi = {\frac 1
  2}\varphi_{ab}\, dx^a\wedge dx^b$, we can consider its expansion
with respect to the basis of 2-forms $\epsilon_a$, namely
$\varphi=\widehat{\varphi}^a\epsilon_a$.  This defines a mapping of
the antisymmetric tensor components $\varphi_{ab}$ into the components
of a vector density of the weight $+1$. Explicitly,
\begin{equation}
  \widehat{\varphi}^1:=\varphi_{23},\quad\widehat{\varphi}^2:=\varphi_{31},
  \quad\widehat{\varphi}^3:=\varphi_{12}.
\end{equation}
In particular, for any two covectors $v_a$ and $u_a$, a ``vector
product'' is defined which maps them into a vector density
$[v\times_\epsilon u]^a$ of weight $+1$, via the relation
$[v\times_\epsilon u]^a\epsilon_a:=v\wedge u$.

This mapping allows to define one more differential operator, this
time acting on 2-forms. At first, one immediately sees that
\begin{equation}
dx^a\wedge \epsilon_b = \delta^a_b\,\epsilon.\label{dxeps}
\end{equation}
Consider now the exterior differential of a 2-form $d\varphi$. Using its
expansion with respect to $\epsilon_a$ and the property (\ref{dxeps}),
one finds $d\varphi=(\partial_a\widehat{\varphi}^a)\epsilon =:\left(
  {\rm div}_{\epsilon}\,\varphi\right)\epsilon$, which defines the
${\rm div}_{\epsilon}$ operator: it maps antisymmetric 2-tensors into
scalar densities of weight $+1$. Explicitly,
\begin{equation} 
{\rm div}_{\epsilon}\,\varphi  :=\partial_1\varphi_{23} +
\partial_2\varphi_{31} +\partial_3\varphi_{12}=
\partial_a\widehat{\varphi}^a.\label{dive}
\end{equation}
One can straightforwardly verify the transformation law under the
change of the local coordinates $x^a\rightarrow x^a(x'{}^b)$:
\begin{equation}
  \left({\rm div}_{\epsilon}\,\varphi\right)\longrightarrow
  \left({\rm div}_{\epsilon}
    \,\varphi\right)' = J^{-1}\left({\rm div}_{\epsilon}\,\varphi\right).
\end{equation}

Equivalently, one can write the operators (\ref{rot}) and (\ref{dive}) with the
help of the totally antisymmetric tensor density: 
\begin{equation}
  \left({\rm curl}_{\epsilon}\,\omega\right)^a =
  \epsilon^{abc}\,\partial_b\,\omega_c,
  \qquad {\rm div}_{\epsilon}\,\varphi = \epsilon^{abc}\,\partial_a\,\varphi_{bc}.
\end{equation}
We collected our main results in Table I. In the next section, we
extend this three-vector formalism to four spacetime dimensions.

%%%%%%%%%%%%%%%%%%%%%
\begin{center}
\noindent{\bf Table I.} Premetric differential operators (we use only
partial derivatives!)\bigskip

%%%%%%%%%%%%%%%%%%%%%%%% BEGIN TABLE %%%%%%%%%%%%%%%%%%%%%%%% 
\noindent
\begin{tabular}{|c|c|c|c|c|}
\hline 
operator    &domain& argument&image &components  \\
\hline 
${\rm grad}$ =  ${\bm \partial}$& scalars& $f$ &  covectors  &   
$ \left( \begin{array}{c}
\partial_1 f\\
\partial_2 f  \\
\partial_3 f\end{array} \right)$  \\
\hline 
${\rm curl}_\epsilon$ =  ${\bm \partial \times}$& covectors &
$\left( \begin{array}{c}
    \omega_1 \\
    \omega_2   \\
    \omega_3 \end{array} \right)$ &    $ \begin{array}{c}
  {\rm vector} \\
  {\rm densities  } \\
  {\rm \!\!weight +1\!\!} \end{array} $&   
\!\!$ \left( \begin{array}{c}
\partial_2 \omega_3-\partial_3 \omega_2\\
\partial_3 \omega_1-\partial_1 \omega_3  \\
\partial_1 \omega_2-\partial_2 \omega_1\end{array} \right)$ \!\! \\
\hline 
${\rm div}_\epsilon$= ${\bm \partial\, \cdot}$& $ \begin{array}{c}
  {\rm  antisym.} \\
  \!\!{\rm 2nd\; rank  }\!\!\\
\text{tensors} %%{\rm weight \,+1} 
\end{array} $&
\!\!$\left( \begin{array}{c}
    {\varphi}_{23}=\hat{\varphi}^{1} \\
    {\varphi}_{31} =\hat{\varphi}^{2}  \\
    {\varphi}_{12} =\hat{\varphi}^{3}\end{array} \right)$\!\!\!\!
& $ \begin{array}{c}
  {\rm scalar} \\
  {\rm densities  } \\
  {\rm\!\! weight+\!1} \!\!\end{array} $  &  
$\partial_1 \hat{\varphi}^1+\partial_2 \hat{\varphi}^2+\partial_3
                           \hat{\varphi}^3
$  \\
\hline
\end{tabular}
%%%%%%%%%%%%%%%%%%%%%%%% END TABLE %%%%%%%%%%%%%%%%%%%%%%%%
\end{center}
\bigskip

%%%%%%%%%%%%%%%%%%%%%%%%%%%%%%%%%%%%%%%%%%%%%%%%%%%%%%%%%%%%%%%%%%%%%
\section{The Maxwell equations in premetric 3-vector
  form}\label{Maxwell}
%%%%%%%%%%%%%%%%%%%%%%%%%%%%%%%%%%%%%%%%%%%%%%%%%%%%%%%%%%%%%%%%%%%%%

In 1907, Minkowski fused the time coordinate $t$ and the space
coordinates $x^a$ into a fundamental structure, the 4-dimensional
spacetime manifold (``world''). Basically, he achieved this by using
Maxwell's field theory of electromagnetism. Let us denote the local
spacetime coordinates by $x^i = (t, x^a)$.

The three building blocks of the classical electrodynamics are the
electromagnetic field strength 2-form $F$, the electromagnetic
excitation 2-form $H$, and the electric current density 3-form $J$.
The Maxwell equations are written in a generally covariant form, which
is valid for all coordinates and for all reference frames:
\begin{eqnarray}
dF &=& 0,\label{maxF0}\\ dH &=& J.\label{maxH0}
\end{eqnarray}
The electromagnetic field strength can be decomposed into the electric
and magnetic fields
\begin{equation}
\bm{E}\qquad {\rm and}\qquad \bm{B}.\label{EB0}
\end{equation}
Their components, ${E}_a$ and ${B}^a$, are identified with the
components of the field strength tensor, $F = {\frac
  12}F_{ij}dx^i\wedge dx^j$, as follows:
\begin{equation} {E}_a = F_{a0},\qquad {B}^1 = F_{23},\qquad {B}^2 =
  F_{31}, \qquad {B}^3 = F_{12}.\label{EBa0}
\end{equation}
Accordingly, the electromagnetic field strength 2-form reads
\begin{equation}
F = -\,dt\wedge {E}_adx^a + {B}^1\,dx^2\wedge dx^3 + 
{B}^2\,dx^3\wedge dx^1 + {B}^3\,dx^1\wedge dx^2.\label{F0}
\end{equation}

It is worthwhile to note that the local spacetime coordinates $x^i =
(t, x^a)$ are absolutely arbitrary---not necessarily the Cartesian
ones. If we change the local coordinates
\begin{equation}\label{coordST}
  x^i\longrightarrow x^i = x^i(x'^j) \qquad \begin{cases}\hspace{7pt} 
    t = t(t',x'^a)\\     x^a = x^a(t', x'^b)\end{cases},
\end{equation}
the electric and magnetic fields transform into
\begin{eqnarray}
  {E}'_a &=&\hspace{9pt} L^b{}_a{E}_b - P_{ab}{B}^b,\label{Enew}\\
  {B}'^a &=& -\,Q^{ab}{E}_b + M^a{}_b{B}^b,\label{Bnew}
\end{eqnarray}
where the transformation matrices read
\begin{eqnarray}\label{LP}
  L^b{}_a = {\frac {\partial x^b}{\partial x'^a}}{\frac {\partial t}
    {\partial t'}} - {\frac {\partial x^b}{\partial t'}}{\frac {\partial
      t}    {\partial x'^a}},\qquad P_{ab} =  \epsilon_{bcd}
  {\frac {\partial x^c}{\partial t'}}{\frac {\partial x^d}{\partial x'^a}},\\
  Q^{ab} = \epsilon^{acd}{\frac {\partial t}{\partial x'^c}}{\frac {\partial x^b}
    {\partial x'^d}},\qquad M^a{}_b = \left(\det {\frac {\partial x^c}{\partial x'^d}}\right) 
  {\frac {\partial x'^a}{\partial x^b}}.\label{QM}
\end{eqnarray}
As we see, for a general spacetime transformation, the components of
electric and magnetic fields are mixed up. In a special case of a {\it
  pure spatial} transformation,
\begin{equation}
t = t',\quad  x^a = x^a(x'^b),\label{coordS}
\end{equation}
the above formulas reduce to
\begin{eqnarray}
{E}'_a &=& {\frac {\partial x^b}{\partial x'^a}}\,{E}_b,\label{Ea}\\
{B}'^a &=& \left(\det {\frac {\partial x^c}{\partial x'^d}}\right) 
{\frac {\partial x'^a}{\partial x^b}}\,{B}^b.\label{Ba}
\end{eqnarray}
In other words, the electric and magnetic fields transform
contragrediently. {}From the 3-dimensional point of view, $\bm{E}$
is a 3-covector, whereas $\bm{B}$ is a 3-vector. This explains the
different position of indices ${E}_a$ vs.  ${B}^a$. Moreover,
according to (\ref{Ba}), the magnetic field is a vector {\it density}
and not a true vector.

In this table, two twisted electric charge and current forms $\rho$
and $j$ represent the source of the electromagnetic field. The
excitation of spacetime is represented by two twisted forms ${\cal D}$
and ${\cal H}$. The force acting on the test particle is determined by
two untwisted forms $E$ and $B$.

It is straightforward to find the homogeneous Maxwell equation in
3-compo\-nents.  Substituting (\ref{F0}) into (\ref{maxF0}), we find
\begin{equation}
\bm{\partial}\times \bm{E} + \bm{\dot{B}} = 0,
\qquad \bm{\partial}\cdot\bm{B}  = 0.\label{maxF1}
\end{equation}
Here the dot denotes the time derivative, $\bm{\dot{}} =\partial_t$,
and the differential operator $\bm{\partial}$ is, in components,
represented by $\{\partial_a\}$. In Cartesian coordinates, this
operator coincides with the nabla $\bm{\nabla}$.  The metric-free
differential operators show up here: we used the alternative boldface
notation $\bm{\partial}\times = {\rm curl}_\epsilon$ and
$\bm{\partial} \,\cdot = {\rm div}_\epsilon$ for the divergence
(\ref{dive}) and the curl (\ref{rot}).

In a similar way, we introduce the magnetic and electric excitations
\begin{equation}
 \bm{\mathcal H}\qquad {\rm and}\qquad\bm{\mathcal D},\label{DH0}
\end{equation}
by identifying their components ${H}_a$ and ${D}^a$ with the
components of the excitation tensor, $H = {\frac 12}H_{ij}dx^i\wedge
dx^j$, as follows:
\begin{equation}
{\mathcal H}_a = H_{0a},\qquad {\mathcal D}^1 = H_{23},\qquad
{\mathcal D}^2 = H_{31},
\qquad {\mathcal D}^3 = H_{12}.\label{HDa0}
\end{equation}
Then the excitation 2-form can be rewritten as
\begin{equation}
  H = dt\wedge {\mathcal H}_adx^a + {\mathcal D}^1\,dx^2\wedge dx^3 + 
  {\mathcal D}^2\,dx^3\wedge dx^1 + {\mathcal D}^3\,dx^1\wedge dx^2.\label{H0}
\end{equation}
Under the change of the spacetime coordinates (\ref{coordST}), we find the 
transformation law
\begin{eqnarray}\label{Hnew}
{\mathcal H}'_a &=& \tau\left(L^b{}_a{\mathcal H}_b + P_{ab}{\mathcal
    D}^b\right),\\
{\mathcal D}'^a &=& \tau\left(Q^{ab}{\mathcal H}_b + M^a{}_b{\mathcal
    D}^b\right).\label{Dnew}
\end{eqnarray}
\begin{center}
\begin{figure}
\includegraphics[width=12cm]{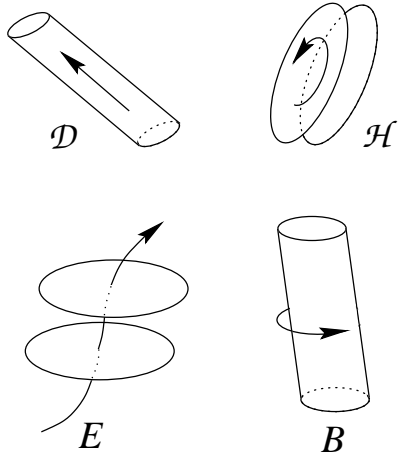}
\caption{Faraday-Schouten pictograms of the electromagnetic
  field, see Schouten \cite{Schouten:1989}.}%\vspace{8pt}
\end{figure}
\end{center}
Here we took into account that $H$ is a twisted 2-form, which is
manifest in the presence of an extra sign factor $\tau = |\det{\frac
  {\partial x^i} {\partial x'^j}}|/\det{\frac {\partial x^i}{\partial
    x'^j}}$. For orientation-preserving coordinate transformations
$\tau = +1$, and $\tau = -1$ in case the orientation is changed. For
the pure spatial transformations, we find
\begin{eqnarray}
  {\mathcal H}'_a &=& {\frac {\left|\,\det {\frac {\partial x^c}
  {\partial x'^d}}\right|}{\det {\frac {\partial x^c}{\partial x'^d}}}}
  \,{\frac {\partial x^b}{\partial x'^a}}\,{\mathcal H}_b,\label{Ea'}\\
  {\mathcal D}'^a &=& \left|\,\det {\frac {\partial x^c}{\partial x'^d}}
  \right| {\frac {\partial x'^a}{\partial x^b}}\,{\mathcal D}^b.\label{Ba*}
\end{eqnarray}
Accordingly, we conclude that $\bm{\mathcal H}$ is a twisted
3-covector, whereas $\bm{\mathcal D}$ is a twisted 3-vector density
with respect to the spatial transformations
(\ref{coordS}). Corresponding attributions can be found in, for
instance, Schouten \cite{Schouten:1989}, Truesdell \& Toupin
\cite{Truesdell:1961}, and in Russer \cite{Russer:2006}, and they are
collected in Table II.
%\newpage

%%%%%%%%%%%%%%%%%%%%%
\begin{center}
\noindent{\bf Table II.} The electric current and the electromagnetic
field\bigskip

%%%%%%%%%%%%%%%%%%%%%%%% BEGIN TABLE %%%%%%%%%%%%%%%%%%%%%%%% 
\noindent
\begin{tabular}{|c|c|c|c|c|c|c|}
\hline 
object & name   & math. & independ. & related & reflec- &absolute \\
 & & object & compon.& to & tion & dimen.\ \\
\hline 
   & electric & & & & & \\
$\rho$ & charge&   twisted &  $\rho$  &   volume &$-\rho$   &  $ q=$ el.\\
& density & 3-form & & &&charge \\
\hline
   & electric & & & & & \\
$j  $  &  current & twisted & $j^a$   & area &$-j  $   &  $q/t$\\
& density & 2-form & & && \\
\hline
${\cal D}$ & electric & twisted & ${\cal D}^1,{\cal D}^2,
{\cal D}^3$  & area & $-{\cal D}$ & $q$\ \\
& excitation & 2-form & & & & \\
\hline
${\cal H}$ & magnetic & twisted & ${\cal H}_1,{\cal H}_2,{\cal H}_3$ 
& line & $-{\cal H}$ & $q/t$\\
& excitation & 1-form & & && \\
\hline
   & electric & & & & & \\
$E$ &  field & untwisted & $E_1,E_2,E_3$ & line & $E$ & $\phi/t$\\
    & strength & 1-form & & & & \\
\hline
   & magnetic  & & & & &\\% $\Phi=$ magnetic\\
$B$ &   field & untwisted & $B^1,B^2,B^3$ & 
area & $B$ & $\phi=$ mag. \\
& strength & 2-form & & & & flux \\
\hline
\end{tabular}

\vspace{0.5cm}

%%%%%%%%%%%%%%%%%%%%%%%% END TABLE %%%%%%%%%%%%%%%%%%%%%%%%
\end{center}
%\bigskip

\begin{center}
\begin{figure}
\includegraphics[width=10cm]{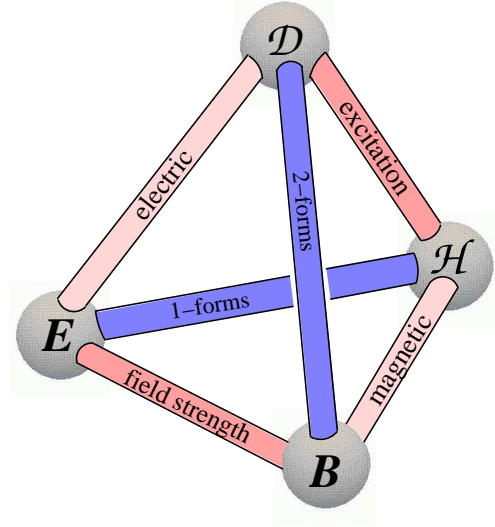}
\caption{Different aspects of the electromagnetic field. The four
  quantities ${\cal D},{\cal H},\,E,B$ jointly constitute the
  electromagnetic field, see Raith's edition of Bergmann-Schaefer
  \cite{Raith:2006}. These four fields are all of a {\it microscopic}
  nature.}
\end{figure}
\end{center}
Finally, we define the electric current density $\bm{J}$ and the electric 
charge density $\rho$ by identifying the components of the current 3-form 
$J = {\frac 16}J_{ijk}dx^i\wedge dx^j\wedge dx^k$:
\begin{equation}
{J}^1 = -\,J_{023},\qquad {J}^2 = -\,J_{031},\qquad 
{J}^3 = -\,J_{012},\qquad \rho = J_{123}.\label{Ja0}
\end{equation}
The current 3-form then reads
\begin{equation}\label{J0}
J = \rho\,dx^1\wedge dx^2\wedge dx^3 - {J}^1dt\wedge dx^2\wedge dx^3 -
{J}^2dt\wedge dx^3\wedge dx^1 - {J}^3dt\wedge dx^1\wedge dx^2.
\end{equation}
It is straightforward to derive the corresponding transformation law, taking 
into account that $J$ is the twisted 3-form:
\begin{eqnarray}
\rho' &=& \left|\,\det {\frac {\partial x^i}{\partial x'^j}}\right| 
\left({\frac {\partial t'}{\partial t}}\rho + {\frac {\partial t'}
{\partial x^b}}{J}^b\right),\label{rnew}\\
{J}'^a &=& \left|\,\det {\frac {\partial x^i}{\partial x'^j}}\right| 
\left({\frac {\partial x'^a}{\partial t}}\rho + {\frac {\partial x'^a}
{\partial x^b}}{J}^b\right).\label{Jnew}
\end{eqnarray}
Note that here we have the determinant of the $4\times 4$ Jacobi
matrix ${\frac {\partial x^i}{\partial x'^j}}$ of the spacetime
coordinate transformation, whereas in (\ref{QM}) and (\ref{Ba}) we
have the determinant of the $3\times 3$ Jacobi matrix ${\frac
  {\partial x^c}{\partial x'^d}}$ of the spatial transformation.
Under the pure spatial transformation (\ref{coordS}),
\begin{eqnarray}
  \rho' &=& \left|\,\det {\frac {\partial x^c}{\partial
        x'^d}}\right|\rho,\label{rr}\\
  {J}'^a &=& \left|\,\det {\frac {\partial x^c}{\partial x'^d}}\right| 
  {\frac {\partial x'^a}{\partial x^b}}\,{J}^b,\label{Ja}
\end{eqnarray}
hence $\rho$ is a twisted 3-scalar density, and $\bm{J}$ is a twisted
3-vector density.  The twisted property is manifest in the modulus of
the determinant in the transformation law, cf. with (\ref{Ba}).

Substituting (\ref{H0}) and (\ref{J0}) into (\ref{maxH0}), we can
recast the inhomogeneous Maxwell equation into
\begin{equation}
\bm{\partial}\times \bm{\mathcal H} - \bm{\dot{\mathcal D}} = \bm{J},
\qquad \bm{\partial}\cdot\bm{\mathcal D} = \rho.\label{maxH1}
\end{equation}

%%%%%%%%%%%%%%%%%%%%%%%%%%%%%%%%%%%%%%%%%%%%%%%%%%%%%%%%%%%%%%%%%%%%%%%%%%%%%%%
\section{Jump conditions derived {}from the premetric Maxwell
  equations{}\footnote{This section is not needed for 
understanding the rationale of the premetric approach. 
Readers who want to concentrate on the premetric and 
epistemological aspects of our article can skip this 
section.}}\label{Jump}
%%%%%%%%%%%%%%%%%%%%%%%%%%%%%%%%%%%%%%%%%%%%%%%%%%%%%%%%%%%%%%%%%%%%%%%%%%%%%%%

In this section, we discuss the jump conditions in premetric
electrodynamics. In the standard metric presentation of the Maxwell
theory, these conditions appear in various situations: (i) On the
timelike boundary between two different types of media; (ii) on the
spacelike Cauchy initial surface; (iii) on the lightlike wave-front
surface.

The jump conditions cannot be treated as independent ingredients of
the theory, as it is presented sometimes in the literature, see for
instance Lakhtakia et al.\ \cite{Lakhtakia:2006, Lakhtakia:2009}.
Quite to the contrary \cite{Sihvola:2008, Obukhov:2009}, these
conditions are straightforward consequences of the field equations.
In the framework of premetric electrodynamics, we are able to
derive unique covariant metric-free jump conditions in a unified
way. When a metric is present, the three different aforementioned
metric-dependent conditions arise as special cases. We present here a
brief discussion, see \cite{Itin:2014wea} for the technical details
and for explicit calculations.

Let us consider an arbitrary surface $S$ defined by the equation
$\varphi(x^i) = 0$.  It is then convenient to derive the covariant
boundary conditions directly {}from the 4-dimensional Maxwell equations
by means of their singular solutions. Observe first that a 2-form $F$,
which is discontinuous on the surface $S$, can be written with the
help of the Heaviside step-function $\theta(\tau)$ as
\begin{equation}\label{F-dist}
F(x^i)={}^{(+)}\!F(x^i)\, \theta(\vp) + {}^{(-)}\!F(x^i)\,\theta(-\vp)
= \begin{cases} {}^{(+)}\!F(x^i),\hspace{5pt} {\rm if}\quad \vp(x^i)> 0,\\
{}^{(-)}\!F(x^i),\hspace{5pt} {\rm if}\quad \vp(x^i) < 0.\end{cases}
\end{equation}
Substituting this expression into the homogeneous Maxwell equation
(\ref{maxF0}), we find
\begin{equation}\label{F-dist-7}
 \big[ F\big] \wedge q = 0\,.
\end{equation}
Here we denoted the jump 2-form by 
\begin{equation}
\big[ F\big] := {}^{(+)}\!F(x^i) - {}^{(-)}\!F(x^i)\,.\label{jumpF}
\end{equation}
We introduced the wave covector (or 1-form)
\begin{equation}
q := d\varphi = q_i\,dx^i,\qquad q_i = \partial_i\varphi.\label{q}
\end{equation}
Making use of the splitting (\ref{EB0})--(\ref{F0}) of the
electromagnetic field strength 2-form $F$ into the electric $\bm{E}$
and magnetic $\bm{B}$ fields, we are able to recast (\ref{F-dist-7})
into a space-time decomposed form
\begin{equation}\label{decomp3}
\big[\bm{B}\big] {\cdot} \bm{q}=0\,, \qquad \big[\bm{E}\big]{\times}\bm{q}
- \big[\bm{B}\big]q_{0}=0\,.
\end{equation}
Here we use the obvious notation $q_i = \{q_0, \bm{q}\}$; the central
dot in (\ref{decomp3}) denotes the contraction of a vector with a
covector ({\it not} the metric-dependent scalar product).

In order to deal with inhomogeneous Maxwell equation (\ref{maxH0}), we
need a description of a current that is localized on the surface
$S$. For such a singular current, we use a representation
\begin{equation}\label{delta-coord-1}
^{(S)}\!J = \d(\vp) j\wedge q\,,
\end{equation}
with the regular 2-form $j = {\frac 12}j_{ij}dx^i\wedge dx^j$.  This
definition guarantees explicitly the principal properties of the
surface current:
\begin{itemize}
\item[(i)] On a 4-dimensional manifold, the surface current is given
  as a 3-form. Thus, it can be directly substituted into the
  inhomogeneous field equation (\ref{maxH0}).

\item[(ii)] The current is localized on the surface $S = \left\{x^i
    \,|\,\vp(x^i) =0 \right\}$ itself and vanishes outside of it. This
  property is provided by the factor $\d(\vp)$.

\item[(iii)] Due to the factor $q$, the current is tangential to the
  surface. Indeed, in view of (\ref{q}), $^{(S)}\!J\wedge d\vp = 0$.

\item[(iv)] The 2-form $j$ and  the 3-form $^{(S)}\!J$ have the same 
absolute dimension of an electric charge.

\item[(v)] The surface current is preserved, even if we choose, for the
  same surface, a different boundary function $\vp(x^i)$.
\end{itemize}

Similarly to (\ref{F-dist}), we use the step-function representation
for the electromagnetic excitation $H$ and introduce the jump 2-form
\begin{equation}
\big[ H\big] := {}^{(+)}\!H(x^i) - {}^{(-)}\!H(x^i)\,.\label{jumpH}
\end{equation}
As a result, in the inhomogeneous Maxwell equation $dH=J$, the current
$J$ is the sum of an ordinary bulk current and the singular
surface current (\ref{delta-coord-1}). As a consequence,
\begin{equation}\label{inh-cond1}
\left(\big[ H\big] - j\right) \wedge q = 0\,.
\end{equation}

In complete analogy with the decomposition of the excitation
(\ref{DH0})--(\ref{H0}), we split the 2-form $j = {\frac
  12}j_{ij}dx^i\wedge dx^j$ into the two pieces $\bm{j}^{\rm s}$ and
$\bm{\sigma}$. We identify the components as follows:
\begin{equation}
j^{\rm s}_a = j_{0a},\qquad \sigma^1 = j_{23},\qquad \sigma^2 = j_{31},
\qquad \sigma^3 = j_{12}\,.\label{JS0}
\end{equation}
Then the current reads,
\begin{equation}
  j = dt\wedge j^{\rm s}_adx^a + \sigma^1\,dx^2\wedge dx^3 +
  \sigma^2\,dx^3\wedge dx^1  + \sigma^3\,dx^1\wedge dx^2.\label{J0'}
\end{equation}

The equation (\ref{inh-cond1}) can then be rewritten in the space-time
decomposed form
\begin{equation}\label{decomp11}
  \big[\bm{\mathcal D}\big] \cdot\bm{q} = \sigma\,, \qquad
  \big[\bm{\mathcal H}
  \big]\times\bm{q} + \big[\bm{\mathcal D}\big]q_{0} = \bm{\mathcal K}\,.
\end{equation}
Here the surface charge and the surface current densities are defined by 
\begin{equation}\label{decomp12}
  \sigma = \bm{\sigma}\cdot\bm{q}\,, \qquad \bm{\mathcal K} = \bm{\sigma}\,q_0 
  + \bm{j}^{\rm s}\times\bm{q}.
\end{equation}
The system (\ref{decomp3}) and (\ref{decomp11}) expresses the jump
conditions for an arbitrary 3-dimensional surface $S$ in a
4-dimensional spacetime. Generally, it represents the condition for a
moving surface, whereas the static case arises for $q_0=0$.

The same conditions are applicable for the wave-front surface. Then
the surface charge and the current are zero, and we obtain the two
premetric conditions for wave-fronts,
\begin{equation}\label{FHdist}
\big[F\big]\wedge q=0\,, \qquad \big[H\big]\wedge q=0\,.
\end{equation}
These conditions are of central importance. They can be used as a
starting point for deriving the dispersion relation for
electromagnetic waves. Incidentally, one can equivalently write the
decomposed jump conditions (\ref{decomp3}) and (\ref{decomp11}) in
terms of components, if needed.

%%%%%%%%%%%%%%%%%%%%%%%%%%%%%%%%%%%%%%%%%%%%%%%%%%%%%%%%%%%%%%%%%%%%%%%
\section{Constitutive law}\label{Constlaw}
%%%%%%%%%%%%%%%%%%%%%%%%%%%%%%%%%%%%%%%%%%%%%%%%%%%%%%%%%%%%%%%%%%%%%%%

The premetric Maxwell equations (\ref{maxF1}) and (\ref{maxH1}) are
generally covariant and valid in all coordinates and reference
systems. Moreover, the spacetime geometry and the metric is not
specified. The geometrical structure of spacetime enters only the {\it
  constitutive law} that relates the excitation $H$ with the field
strength $F$. The constitutive relation depends on the specific
dynamical contents of the electrodynamic theory.

In Maxwell-Lorentz electrodynamics, the constitutive law in vacuum
reads
\begin{equation}\label{const}
  H = \lambda_0\,^\star\!F,\qquad \lambda_0 = \sqrt{\frac
    {\varepsilon_0}{\mu_0}}.
\end{equation}
Here $\varepsilon_0$ and $\mu_0$ are the electric and magnetic
constants and $\lambda_0$ determines the vacuum admittance of about
$1/(377 \text{ ohm})$. Note that $\lambda_0$, in contrast to
$\varepsilon_0$ and $\mu_0$, is a 4-dimensional diffeomorphism
invariant {\it scalar}. The star $^\star$ denotes the Hodge duality
operator determined by the 4-dimensional spacetime metric.

Let us split the local and linear constitutive law (\ref{const}) into
space and time. Then the excitation $H$, according to (\ref{H0}),
decomposes into the vector density $\bm{\mathcal D}$ with components
${\mathcal D}^{a}$ (upper index!) and a covector $\bm{\mathcal H}$
with components ${\mathcal H}_{a}$ (lower index); here
$a,b=1,2,3$. Similarly, the field strength $F$, according to
(\ref{F0}), decomposes into a covector $\bm E$ with components $E_b$
and a vector density $\bm{B}$ with components $B^b$, respectively. If
the 3-dimensional positive definite metric has components $g^{ab}$,
the Maxwell-Lorentz spacetime relation (\ref{const}) becomes
($g:=\text{det}\,g_{cd}$)
\begin{equation}\label{constM}
  {\mathcal D}^{\,a}=\varepsilon_0\,\sqrt{g}g^{ab}\,E_b\,,\qquad  
  {\mathcal H}_{\,a}=\frac{1}{\mu_0}\,\frac{1}{\sqrt{g}}g_{ab}\,B^b\,.
\end{equation}
In Cartesian coordinates, $\sqrt{g}g^{ab} \stackrel{*}{=} \delta^{ab}$, 
$\frac{1}{\sqrt{g}}g_{ab} \stackrel{*}{=} \delta_{ab}$.
%\newpage
If we introduce the 3-tensor density $\mathfrak{g} =
\sqrt{g}g^{ab}$ and its inverse $\mathfrak{g}^{-1}=\frac{ 1}{\sqrt{g}}
g_{ab}$, we can collecting our results in Table III:
\newpage
\begin{center}
  \noindent{\bf Table III.} The premetric Maxwell equations with the
  Maxwell-Lorentz spacetime relation for vacuum\bigskip

%%%%%%%%%%%%%%%%%%%%%%%% BEGIN TABLE %%%%%%%%%%%%%%%%%%%%%%%%
\noindent
\begin{tabular}{|l|c|}
  \hline
  Physics law   & Math. expression  \\ \hline\hline
  Amp\`ere-Maxwell law&  
  $\bm{\partial}\times \bm{\mathcal H} - \bm{\dot{\mathcal D}} =
  \bm{J}$\\ 
  \hline Coulomb-Gauss law&\hspace{25pt}
  $\bm{\partial}\cdot\bm{\mathcal D} = \rho$ \\ \hline
  Faraday induction law& $\bm{\partial}\times \bm{E} + \bm{\dot{B}} =
  0$ \\ \hline
  conserved magnetic flux&\hspace{25pt} $\bm{\partial}\cdot\bm{B}  =
  0$  \\ \hline\hline
  permittivity of vacuum&  $  \bm{\mathcal D}=\hspace{5pt}\varepsilon_0\,
  \hspace{5pt}\mathfrak{g}\,\hspace{5pt}\bm{E}$\\ \hline
  permeability of vacuum& \hspace{5pt}$ \bm{\mathcal H}=\mu_0^{-1}\,
  \mathfrak{g}^{-1}\,\bm{B}$ \\ \hline
\end{tabular}
%%%%%%%%%%%%%%%%%%%%%%%% END TABLE %%%%%%%%%%%%%%%%%%%%%%%%
\end{center}
%%%%%%%%%%%%%%%%%%%%%%%%%%%%%%%%%%%%%%%%%%%%%%%%%%%%%%%%%%%%%%%%%%%%%%%%%

Our scheme is consistent with the Tonti diagram [ELE3] on
electromagnetism in vector notation, see Tonti
\cite[p.\,412]{Tonti:2013}. However, our Maxwell equations are genuinely
metric-free. Gronwald et al.\ \cite{Gronwald:2005tv} studied this
formalism and its relation to gauge field theories.

%%%%%%%%%%%%%%%%%%%%%%%%%%%%%%%%%%%%%%%%%%%%%%%%%%%%%%%%%%%%%%%%%%%%%%%%%
\section{A legacy of Kottler: premetric fundamental 
physical constants versus the speed of light}\label{constants}
%%%%%%%%%%%%%%%%%%%%%%%%%%%%%%%%%%%%%%%%%%%%%%%%%%%%%%%%%%%%%%%%%%%%%%%%

\subsection{Vacuum admittance $\lambda_0$, vacuum speed of light
  $c$}\label{61}

The Maxwell-Lorentz law invites us to look at the physical dimensions
involved in this expression. A physical quantity $Q$ is by definition
$Q$ = \{nume\-rical value\} $\times$ [unit], see Wallot
\cite{Wallot:1953}, Stille \cite{Stille}, and Flowers
\cite{Flowers:2004}. A placeholder for the unit is the physical
dimension [Q], which leaves the choice of the system of units open,
even though we will use here for convenience SI (speak: ``the
International System of Units''). In SI, a law in physics can be
expressed as a {\it quantity equation,} that is, as an equation
relating different physical quantities.\footnote{In contrast, one
  could use purely {\it numerical} equations in physics; they are only
  valid in a predetermined system of units, see Jackson \cite{Jack}
  who oscillates in his whole book between numerical equations for SI
  and those for the Gaussian system. However, quantity equations, like
  those in our Table III, are universally valid for an arbitrary
  choice of a unit system. If a fundamental physical equation, which
  is {\it not} related to spherical symmetry, a ``$\pi$'' appears, you
  can be sure that it is a numerical equation valid only in a specific
  unit system. For a collection of numerous different unit systems in
  electrodynamics, see Carron \cite{Carron:2015rea}.}

{}From elementary electrostatics we know that the physical dimensions
of the electric excitation and the electric field strength are
\begin{align}\label{DE}
[\mathcal{D}^a]&=\frac{\text{\it el.charge}}{\text{\it
    area}}=\frac{q}{{l}^{2}}
\stackrel{\text{SI}}{=}\frac{\text{coulomb}}{\text{meter}^2}=
\frac{\rm C}{{\rm m}^{2}}\qquad\text{and}\\
[E_b]&=\frac{\text{\it voltage}}{\text{\it length}}=\frac{\text{\it mag.flux}}
{\text{\it length}\times \text{\it time}}
=\frac{{\phi}}{lt}\stackrel{\text{SI}}{=}\frac{\text{weber}}
{\text{{meter}$\times$ second}}=\frac{\rm W\!b}{\rm m\,s}\,,
\end{align}
respectively. For determining the physical dimension of
$\varepsilon_0$, we turn to orthonormal frames, since we now must have
a metric $g^{ab}$, since $\mathcal{D}^a$ carries an upper index and
$E_b$ a lower one; then, we can compute $E^a:=g^{ab}E_b$ and are able,
since $[g^{ab}]=1$, to compare it with $\mathcal{D}^a$. We find
\begin{equation}\label{epsilon}
[\varepsilon_0]=\frac{[\mathcal{D}^a]}{[E_b]}=\frac{qt}{\phi l}
=\frac{[\lambda_0]}{v}\stackrel{\text{SI}}{=}\frac{\text{second}}{\text{ohm}
  \times \text{meter}}=\frac{\rm s}{\Omega\, \rm m}\,.
\end{equation}
The {\it vacuum admittance} $\lambda_0$, with
$[\lambda_0]:=q/\phi\stackrel{\text{SI}}{=}$\,1/ohm, has the dimension
of a reciprocal resistance.

Similarly, in magnetostatics we find for magnetic excitation and field
strength, respectively,
\begin{align}\label{HB}
[\mathcal{H}_a]=\frac{\text{\it current}}{\text{\it length}}=
\frac{q}{lt}\stackrel{\text{SI}}{=}\frac{\rm C}{\rm
  m\,s}\,\quad\text{and}\quad [B^b]=\frac{\text{\it mag.flux}}{\text{\it
    area}} =\frac{\phi}{l^2}\stackrel{\text{SI}}{=}\frac{\rm
  W\!b}{{\rm m}^2}\,.
\end{align}
Thus, the dimension of the magnetic constant turns out to be 
\begin{equation}\label{mu}
  [\mu_0]=\frac{[B^b]}{[\mathcal{H}_a]}=\frac{1}{[\lambda_0]
    v}\stackrel{\text{SI}}{=}\frac{\Omega\,\rm s}{\rm m}\,.
\end{equation}

It is surprising, the innocently looking non-relativistic analysis
within electro- and magnetostatics led us to the physical dimensions
$[\varepsilon_0]$ and $[\mu_0]$ of $\varepsilon_0$ and $\mu_0$, see
(\ref{epsilon}) and (\ref{mu}), respectively. They contain a clear
message: the square-roots of their quotient and of their product carry
a simple physical interpretation:
\begin{equation}\label{epsilonmu}
  \lambda_0:=\sqrt{\frac{\varepsilon_0}{\mu_0}}\stackrel{\text{SI}}{\approx}
  \frac{1}{377 \,\Omega}\,,\qquad c:=\frac{1}{\sqrt{\varepsilon_0\mu_0}}
  \stackrel{\text{SI}}{\approx}3\times 10^8\, \frac{\rm m}{\rm s}\,.
\end{equation}
The electrodynamic vacuum is characterized by the vacuum resistance
(impe\-dance) and by the vacuum speed of light.\footnote{The speed of
  light $c$, or rather $\sqrt{2}\,c$---because of different units and
  conventions---had already been measured {\it electromagnetically} by
  Weber \& Kohlrausch in 1855, before Maxwell's equations were derived
  in 1865. In fact, Maxwell used this result in his considerations in
  an essential way. It seems, however, that Weber \& Kohlrausch had
  not been aware that the constant they determined is related to the
  speed of light.} These are the two moduli of the electrodynamic
``{\ae}ther,'' that is, the vacuum admittance $\lambda_0$ and the
vacuum speed of light $c$ are two ``moduli'' specifying the
electromagnetic properties of the vacuum.

\subsection{Absolute dimension of a physical quantity}\label{62}
The {\it absolute dimension} of a quantity is the physical dimension
of its corresponding exterior differential form. As exterior form,
these quantities are scalars under diffeormorphisms and under frame
transformations. Therefore, the notion of an absolute dimension has a
covariant meaning and is, thus, of fundamental importance. 

Since the 3d Levi-Civita symbol $\epsilon_{abc}=\pm1,0$ represents
sheer numbers, we have, using (\ref{DE}) and (\ref{HB}), for the
absolute dimensions of the excitations $\mathcal{D}$ and
$\mathcal{H}$,
\begin{equation}\label{DH*}
[\mathcal{D}]=\frac 12\,\varepsilon_{abc}[\mathcal{D}^a\,dx^b\wedge
dx^c]=q\quad\text{and}\quad [\mathcal{H}]=
\mathcal{H}_a\,dx^a=qt^{-1}\,.
\end{equation}
Accordingly, we arrive for the four-dimensional excitation $H$ simply
at the electric charge $q$:
\begin{equation}\label{H*}
  [H]=[dt\wedge\mathcal{H}]+[\mathcal{D}]=\text{\it charge}=
  q\stackrel{\text{SI}}{=}{\rm C}=\text{coulomb}\,.
\end{equation}
For the field strength $E$ and $B$ we have analogously
\begin{equation}\label{absoloute}
  [E]=[E_a\wedge dx^a] =\phi t^{-1}\,\quad\text{and}\quad [B]=\frac 12
  \epsilon_{abc}[B^a dx^b\wedge dx^c]=\phi
\end{equation}
or, for the four-dimensional field strength $F$,
\begin{equation}\label{F*}
  [F]=-[dt\wedge E]+[B]=\text{\it magnetic flux}=\phi
  \stackrel{\text{SI}}{=}{\rm W\!b}
  =\text{weber}\,.
\end{equation}

If we take the theory of the absolute dimensions, with $[H]=q$ and
$[F]=\phi$, to its logical conclusion, we can extract information by
computing the quotient and the product of them, with $\frak{h}$ as the
dimension of an action,\footnote{The 4d scalars, electric charge $q$,
  magnetic flux $\phi$, and impedance $1/\lambda_0$, all have
  rightfully own names in SI, namely $\rm C,\,W\!b,\,\text{and
  }\Omega$, respectively. They really do merit own names. However, the
  action has none. We suggest $\rm P\hspace{-1pt}\ell = planck$.
  Then, $\rm planck = coulomb \times weber= joule\times second$, or,
  $\rm P\hspace{-1pt}\ell =CW\!b=AsV\!s=V\!As^2=Ws^2=J\,s$.}
\begin{equation}\label{planck}
  \frac{[H]}{[F]}=\frac{q}{\phi}=[\lambda_0]\stackrel{\text{SI}}{=}\frac{1}
  {\Omega}\,, \; [H]\wedge[F]=q\,\phi=\text{\it action}=
  \frak{h}\stackrel{\text{SI}}{=}{\rm J\,s}=\text{joule}\times\text{second}\,.
\end{equation}

Accordingly, the absolute dimensions of the excitation $H$ and the
field strength $F$ led us to the 4d scalars {\it electric charge} $q$
and {\it magnetic flux} $\phi$, their quotient to an {\it admittance}
with dimension $[\lambda_0]=q/\phi$ and their product to an {\it
  action} with dimension $\frak{h}=q\phi$. In SI, we have,
respectively, $\rm C$, $\rm W\!b$, $1/\Omega$, and $\rm Js$. These
notions, as well as their SI expressions, are 4d diffeomorphism
invariant scalars. This is the message of {\it classical}
electrodynamics, facts, which are hardly mentioned in the textbook
literature. Note in particular that the magnetic flux and the action
feature in a purely classical situation, no quantum aspects were
involved. Particularly noteworthy is that the speed of light does {\it
  not} occur here. This is clear since we know that the speed of light
changes at the presence of a gravitational field---in contrast to
electric charge and magnetic flux, which are unaffected by gravity. In
this sense, $c$ must not enter here---and it does not!

These facts haven been already noted by Fleischmann
\cite{Fleischmann:1971}, for example: there are premetric
diffeomorphism invariant 4-scalars in physics, let us call them
4d-scalars, namely $q,\,\phi,\,[\lambda_0],\,\frak{h}$, and those
related to the existence of a metric and connected to space and time
differences, with the speed of light, $c$, as an example. We call $c$
as a P-scalar, since $c$ is only a scalar under the Poincar\'e group
of special relativity. We consider the premetric 4d-scalars as more
fundamental than the metric-dependent P-scalar $c$.

The premetric 4d-scalars, which we mentioned here, were extracted {}from
mechanics ($\mathfrak h$) and electrodynamics
($q,\phi,[\lambda_0]=q/\phi$) alone. {}From thermodynamics, the entropy
$S$ should be added to this list of 4d-scalars.

\subsection{Counting procedures, the Josephson constant $K_{\rm J}$
  and the von Klitzing constant $R_{\rm K}$}\label{63}

How can we understand our 4d covariant results {}from a more intuitive
point of view? Well, we just have to look at an axiomatics of
classical electrodynamics \cite{Birkbook}. The inhomogeneous Maxwell
equation (\ref{maxH0}), $dH=J$, results {}from the postulated
conservation of electric charge $Q=\int_3 J$. As is known {}from a proof
of the Stokes theorem, no metric concepts are involved, rather---if
applied to the charge 3-form---the counting of charges, and addition
or subtraction depending on their sign, is all that is required. If we
prescribe an arbitrary three dimensional domain (a 3d ``volume''), we
have just to know how many charges are contained therein. The net
change of the charge in the course of time has to flow through the
2-boundary encircling the 3d domain.

We describe classical electrodynamics. But, of course, in the back of
our mind is the knowledge that in nature there are quantized
elementary charges with electric charge of
$e\stackrel{\text{SI}}{\approx}1.602\times 10^{-19}\,\rm C$. This
makes us sure that the counting process has a solid basis in
experimental physics. In this way, the concept of electric charge
enters electromagnetism.

The homogeneous Maxwell equation (\ref{maxF0}), $dF=0$, emerges {}from a
careful 4d discussion of Faraday's induction law \cite{Birkbook}. It
turns out that this law can be understood as a conservation law of the
magnetic flux $\Phi=\int_2 F$. Please note that this conservation law
refers to the 2-form $F$, whereas usually we relate a conservation law
to a 3-form, see the case of the electric charge. A geometrically
trained physicist has no difficulty to understand this fact, but those
attached to 3d vector calculus may find this counterintuitive and may
go astray to the outdated magnetic charge concept instead.

But how is it with the counting of elementary flux (lines)?
Certainly, in classical theory, there is no discrete flux. But already
Faraday spoke of lines of force, a line being a discrete entity. We
know {}from experimental physics that under certain circumstances,
namely in type II (two) superconductors, magnetic flux is quantized in
terms of flux units (Abrikosov vortices) with
$\Phi_0:=h/(2e)\stackrel{\text{SI}}{\approx}2.068\times 10^{-15}\rm
W\!b$, with $h$ as the Planck constant; the factor $2$ occurs because
a Cooper pair in the superconductor is built up {}from {\it two}
electrons.

Certainly, in nature the magnetic flux is not quantized in
general. Magnetic fluxes of currents in our brain can be appreciably
smaller that the flux quantum. Still, it is clear that the notion of
counting flux lines exists in nature under special circumstances. This
speaks in favor of the reasonableness of the counting procedure also
for the magnetic flux in general. After all, we know that the Faraday
law is very well fulfilled in nature: This is the way the magnetic
flux emerges in electrodynamics.

Having electric charge and magnetic flux now at our disposal and
having recognized that the corresponding counting procedures are based
on the elementary charge $e$ and the flux quantum $h/(2e)$, it is
clear that physically the {\it quantized nature} of the {\it charge}
($e$ is a 4d scalar) and of the {\it action} ($h$ is a 4d
scalar\footnote{Post \cite{Posthbar} argued that the action could be,
  perhaps, a pseudoscalar instead.}) are the raisons d'{\^e}tre for
the possibilities of counting.  Accordingly, the dimensions $q$ of
charge and $\frak{h}$ of action are fundamentally distinguished {}from
other physical dimensions, such as mass $m$, length $l$, and time $t$.

If $q$ and $\mathfrak{h}$ are premetric scalars, the same must be
  true for the expressions
\begin{equation}\label{scalars}
  q^{n_1}\,\frak{h}^{n_2}\,=\,\mbox{4d {\rm scalar}}\,.
\end{equation}
Strictly, by our arguments $n_1$ and $n_2$ are not required to be
integers. However, examples for such dimensionful 4-scalars with
integers are
\begin{equation}\label{examplesSC}
  q\rightarrow\mbox{\rm electric charge}\,,\quad \frac{\frak{h}}{q}\rightarrow
  \mbox{\rm magnetic flux}\,, \quad
  \frac{\frak{h}}{q^2}\rightarrow\mbox{\rm electric resistance}\dots\,.
\end{equation}
And here we only observe $n_1=\pm1,-2$ and $n_2=0,1$.

We discussed these results previously, see \cite{Hehl:2004jn}. If we
pick for $q$ the elementary charge $e$ and for $\frak{h}$ the Planck
constant $h$, we immediately arrive at the the {\it Josephson\/}
\cite{Josephson,Andreone} and the {\it von Klitzing\/}
\cite{Klitzing,Jeckelmann} constants of modern metrology---for
reviews, see
\cite{Flowers,metro1,CODATA:2012,metro2,Haddad:2016}---which provide
highly precise measurements\footnote{In SI: kilo=k=$10^3$,
  mega=M=$10^{6}$, giga=G=$10^9$, tera=T=$10^{12}$, peta=P=$10^{15}$,
  exa=E=$10^{18}$.} of $e$ and $h$,
\begin{equation}
  K_{\rm J} = {\frac {2e} {h}}\stackrel{\rm SI}{\approx} 0.483 \;{\rm PHz/V},
  \qquad R_{{\rm K}} = \frac{h}{e^2}\stackrel{\rm SI}{\approx}
  25.813\;\rm k\Omega\,;\label{KR}
\end{equation}
for new experiments to the quantum Hall effect see Weiss \& von
Klitzing \cite{Weis:2011} and for the theory Bieri \& Fr\"ohlich
\cite{Bieri:2010za}, and for quantum metrology generally, G\"obel \&
Siegner \cite{Gobel:2015}. The Planck units of 1899 are discussed by
Kiefer \cite{Kiefer}. Note that the reciprocal of the Josephson
constant $ K_{\rm J}$ has the absolute dimension $\phi$ of a magnetic
flux, the von Klitzing constant $R_{{\rm K}}$, of course, that of an
electric resistance (impedance). Incidentally, the premetric nature of
$R_{\rm K}$ has helped us to predict \cite{QHEgrav} that the quantum
Hall effect is {\it not} influenced by gravity (that is, the Hall
resistance is independent of the gravitational field).

So far, our dimensional analysis in the subsections \ref{62} and
\ref{63} did not make use of the metric.  Moreover, it is generally
covariant and as such valid in particular in general relativity as
well as in special relativity. And, on top of that, our considerations
do not depend on any particular choice of the system of physical
units. Whatever your favorite system of units may be, our results will
apply to it. In short: Our dimensional analysis so far is {\it
  premetric, generally covariant, and valid for any system of
  units}---and it led straightforwardly to the 4d scalars $q,\phi,
[\lambda_0]=q/\phi, \frak{h}$.

\subsection{The vacuum speed of light at last}\label{64}

But why did the speed of light {\it not} occur in our analysis of the
absolute dimensions? Simply because the measurement of a velocity
requires a spatial and a temporal distance concept, that is, the
additional existence of a metric. In connection with measuring a
velocity, besides counting, distances are relevant.

For the analysis of the premetric structure of the Maxwell equations,
the counting procedure was all we needed. And it led to the premetric
funda\-mental constants with the dimensions of $q,\phi, [\lambda_0],
\frak{h}$. More is required for the understanding of the constitutive
law of Maxwell-Lorentz, see Table III.

The Maxwell-Lorentz law describes the electromagnetic properties of
the vacuum, as we saw in Sec.\,\ref{61}. Clearly, this vacuum is
assumed to be homogeneous and isotropic. Consequently, the Euclidean
group, as a semidirect product of the translations $T(3)$ with the
rotations $SO(3)$, underlies the geometry of 3d space. In the 4d {\it
  Minkowski spacetime} of special relativity, the group of motion is
the Poincar\'e group, in which $T(3)$ is generalized to $T(4)$ and
$SO(3)$ to the Lorentz group $SO(1,3)$. The existence of a unique
isotropic speed of light attests to the isotropy of spacetime. In an
anisotropic spacetime,\footnote{Incidentally, attempts were made to
  develop a corresponding generalization of electrodynamics and of
  special relativity on the basis of the Finsler geometry
  \cite{Bogoslovsky:1&2,Bogoslovsky:3}.}  the velocity of light would
depend on the direction of propagation. Accordingly, we will be able
to find an unique speed of light in the homogeneous and isotropic
Minkowski spacetime.

Let us recall of how anisotropy of light propagation can be
achieved. We recall the constitutive vacuum law in electrostatics: $
\bm{\mathcal D}=\varepsilon_0\,\mathfrak{g}\,\bm{E}$, or, in
components, $ {\mathcal D}^{\,a}=\varepsilon_0\,\sqrt{g}g^{ab}\,E_b$,
see Table III. For a local and linear dielectric material, which is
additionally homogeneous and isotropic, this law generalizes to
$\bm{\mathcal D}=\varepsilon\varepsilon_0\,\mathfrak{g}\,\bm{E}$, with
the (dimensionless) permittivity $\varepsilon$. If the dielectric is
anisotropic, as was known already to Maxwell, $\varepsilon$ becomes a
tensor $\varepsilon^a{}_b$ such that we find
\begin{equation}\label{anisotropic} 
{\mathcal D}^{\,a}=\varepsilon^a{}_c\,\varepsilon_0\,\sqrt{g}
g^{cb}\,E_b=(\varepsilon_0\sqrt{g}\varepsilon^{ab})E_b\,.
\end{equation}
The expression within the parentheses we call the permittivity tensor
density. Note that the 3d metric and its scalar density $\sqrt{g}$ is
included in this expression.

Analogous consideration, also already performed by Maxwell, can be
made in magnetostatics with the permeability tensor density,
$B^a=(\mu_0\sqrt{g}\mu^{ab})$ $ H_b$. We can additionally allow for
cross-terms between the magnetic and the electric effects, which occur
in so-called {\it magneto-electric} media, see O'Dell
\cite{O'Dell:1970}. Then the four-dimensional generalization of our
anisotropic law, generalizing the isotropic law $ H =
\lambda_0\hspace{1pt}^\star\!F$, can be written as
\begin{equation}\label{const&}
  {H}_{ij} = \frac 12\kappa_{ij}{}^{kl}F_{kl}\,,\qquad\text{with}\qquad
\kappa_{ij}{}^{kl}=-\kappa_{ji}{}^{kl}=-\kappa_{ij}{}^{lk}
\end{equation}
as the doubly antisymmetric {\it local and linear response} tensor. It
has 36 independent components and the dimension $[\kappa_{ij}{}^{kl}]$
of an admittance $[\lambda_0]$. Also here, the metric and its scalar
density as well as the electric and the magnetic constants are
included within the response tensor density. In Sec.\,\ref{Linear} we
will turn to a discussion of this law.

For a specific choice of the response tensor $\kappa_{ij}{}^{kl}$, we
recover the Maxwell-Lorentz vacuum case (\ref{const}), namely for
\cite[p.\ 303]{Birkbook}
\begin{equation}\label{vacuum}
  \kappa_{ij}{}^{kl}=\lambda_0\,\epsilon_{ijmn}\,\sqrt{-g}\,g^{mk}g^{nl}\,,
  \qquad \lambda_0=\sqrt{\frac{\varepsilon_0}{\mu_0}}\,.
\end{equation}
Here $\epsilon_{ijkl}=\pm1,0$ is the totally antisymmetric Levi-Civita
tensor density and $g^{kl}$ is the metric of the Minkwoski
spacetime. Then, of course, the vacuum response tensor is determined,
besides by the vacuum admittance $\lambda_0$, only by the components
$g^{kl}$ of the metric of the Minkwoski spacetime.

We substitute the vacuum constitutive law (\ref{const&}) cum
(\ref{vacuum}) into the premetric Maxwell equations, which, in
components, read (compare with Einstein \cite{Einstein:1916})
\begin{equation}\label{bothMax}
\partial_{[i}H_{jk]}=J_{ijk}\,,\qquad\partial_{[i}F_{jk]}=0\,.
\end{equation}
%%%%%%%%%%%
Subsequently, we can derive the wave equation and discover that the
propagation of electromagnetic waves is ruled by the speed of light
$c:=1/\sqrt{\varepsilon_0\mu_0}$.

This is how the speed of light enters the scene. Not before the vacuum
law (\ref{vacuum}) is assumed. The vacuum admittance $\lambda_0$, a
premetric concept, is already incorporated in (\ref{const&}) in the
form of the physical dimension $[\kappa_{ij}{}^{kl}]=[\lambda_0]$ of
$\kappa_{ij}{}^{kl}$. The speed of light requires for its emergence
the Minkowski metric in (\ref{vacuum}). The speed of light $c$---in
contrast to the vacuum admittance $\lambda_0$---is a metric-dependent
quantity willy nilly. This is also one of the lessons one learns in
looking at electrodynamics with the eyes of Kottler. 

%%%%%%%%%%%%%%%%%%%%%%%%%Last piece of Sec.\,6. %%%%%%%%%%%%%%%%%%%%%%%%%%%%%

Let us recall how Einstein constructed SR (special relativity),
neglecting gravity, in 1905, for the time being. In ``The Meaning of
Relativity'' \cite{Meaning}, we read: {\it ...the Maxwell-Lorentz
  equations have proved their validity in the treatment of optical
  problems in moving bodies. No other theory has satisfactorily
  explained the facts of aberration, the propagation of light in
  moving bodies (Fizeau), and phenomena observed in double stars (De
  Sitter). The consequence of the Maxwell-Lorentz equations that in a
  vacuum light is propagated with the velocity $c$ {\em (``principle
    of the constancy of the velocity of light''),}\footnote{Note that
    the phrase in the parentheses, which we printed in roman letters,
    was {\it not} translated into English, even though it appeared in
    the German original. Incidentally, in English, one often talks
    about the {\it speed} of light, referring to its modulus.} at
  least with respect to a definite inertial system $K$, must therefore
  be regarded as proved. According to the principle of special
  relativity, we must also assume the truth of this principle for
  every other inertial system.} Thus, SR, according to Einstein, is
based (i) on the special relativity principle (equivalence of all
inertial frames of reference) and (ii) on the principle of the
constancy of the speed of light. Because of (ii), the speed of light
is, in SR, a scalar under Poincar\'e transformations (a $P$-scalar) by
construction. Needless to say that Einstein's conception was highly
successful and, in 1945, the Alamogordo bomb was the final and widely
visible proof of the validity special relativity.

In GR (general relativity), one postulates the existence of a constant
$c_0 = 299\, 792. 458\,$km\,s$^{-1}$. However, $c_0$ is {\it not} an
actual speed of light in the presence of the gravitational
field. Recall, that in a medium with the refractive index $n$, the
light propagates with the velocity $c_0/n$, when the medium is at rest
and we neglect gravity. In inhomogeneous media, the refractive index
is a function of the coordinates, $n = n(\bm{x})$. The gravitational
field is a special type of an ``inhomogeneous medium'' with the
permittivity, permeability, and the magneto-electric moduli
constructed {}from the components of the spacetime metric $g$. The
latter contains information about the true gravitational field as well
as about the inertial field arising due to the motion of the reference
frame. Accordingly, in GR the actual speed of light $c$ depends on
this field $g$; symbolically, $c=c(g)$.

This fact is well established {}from a great number of observations. In
particular, we see that the light of distant stars is deflected by the
gravitational field $g$ of the Sun, and space missions reveal numerous
time delay effects for light and radio signals. When gravity is
present, which is always the case due to its universality, $c_0$
cannot be measured optically as speed of light in GR. For the
measurement of a velocity, one needs two different points in
spacetime; and always there are tidal gravitational forces between
those 2 points, which cannot be transformed to zero even in a freely
falling laboratory. The speed of light is a $P$-scalar, but it is not
a four-dimensional spacetime scalar, it is not diffeomorphism
invariant. The $c_0$ is by assumption a four-dimensional scalar, but
it is no longer the speed of light. Still, in all textbooks one reads
that $c_0$, confusingly, is called the speed of light.

In other words, the principle of the constancy of the velocity of
light loses its general validity. Einstein \cite{Einstein:1907} stated
that (our translation) {\it The principle of the constancy of the
velocity of light allows also here [when gravity is present] the
definition of simultaneity, provided one restricts oneself to very
small light paths.} That is, the principle of the constancy of the
velocity of light is only applicable in a certain limiting case.

One could argue that the scalar nature under {\it diffeomorphism} of
$c_0$ is a new principle of GR, which generalizes the principle of the
constancy of of the speed of light $c$ in SR. For a discussion of
$c_0$ and the cosmological constant, see Dadhich
\cite{Dadhich:2011gx} and for Einstein's introduction of the
cosmological constant Sauer \cite{Sauer:2012}.

In a recent paper, Braun, Schneiter, and Fischer \cite{Braun:2015twa}
discuss how precisely one can measure the speed of light. They find
quantum effects that limit the corresponding precision in
principle. Here we similarly argue that we have already classical
effects of gravity, which likewise limit the precision of the
determination of the speed of light.

As we can see, the division of universal constants in four-dimensional 
(diffeomorphism) scalars and $P$-scalars marks a decisive qualitative 
difference between the universal constants.

In modern metrology \cite{CODATA:2012,SI:2006} the value of the
constant $c_0 = 299\, 792.  458\,$ km\,s$^{-1}$ is {\it exact}, and the
role of $c_0$ is reduced to a conversion factor that is used to define
an SI unit of length: the meter is introduced as the length which the
light passes in vacuum during a time that equals 1/299\,792\,458 of a
second.  The possibility of a similar definition of a kilogram, by
making use of the fundamental physical constants, is now widely
discussed, see Haddad et al.\ \cite{Haddad:2016}.

%%%%%%%%%%%%%%%%%%%%%%%%%End of Sec.\,6. %%%%%%%%%%%%%%%%%%%%%%%%%%%%%

%%%%%%%%%%%%%%%%%%%%%%%%%%%%%%%%%%%%%%%%%%%%%%%%%%%%%%%%%%%%%%%%%%%%%%%%%
\section{Kottler's program for Newton's gravity}\label{Newton}
%%%%%%%%%%%%%%%%%%%%%%%%%%%%%%%%%%%%%%%%%%%%%%%%%%%%%%%%%%%%%%%%%%%%%%%%
%{{\bf To 1.} Newton-Poisson law, 3d, gravito-statics}

When in electrodynamics the electric charge density $\rho_{\rm el}$
does not depend on time and neither electric currents nor magnetic
fields are present, the Maxwell equations (\ref{maxH1}) reduce to the
electrostatic law: $\bm{\partial}\cdot\bm{\mathcal D} = \rho_{\rm
  el}$. The electric charge density $\rho_{\rm el}$, a scalar density,
creates the electric excitation $\bm{\mathcal D}$.

Kottler noticed that Newton's gravity can be treated similarly to
electrostatics.  Now, instead of the electric charge density, the mass
density $\rho_{{\rm m}}$, also a scalar density, induces the
gravitational excitation $\bm{\mathcal D}_{\text{gr}}$ with physical
dimension $[\bm{\mathcal D}_{\text{gr}}]=m/ l ^2$:
\begin{equation}\label{NGauss}
\bm{\partial}\cdot\bm{\mathcal D}_{\text{gr}} = -\,\rho_{\rm m}.
\end{equation}
The minus sign appears in gravity because of the gravitational
attraction of masses as opposed to the repulsion of electrically
equally charged particles.
\begin{center}
\begin{figure}
\includegraphics[width=9cm]{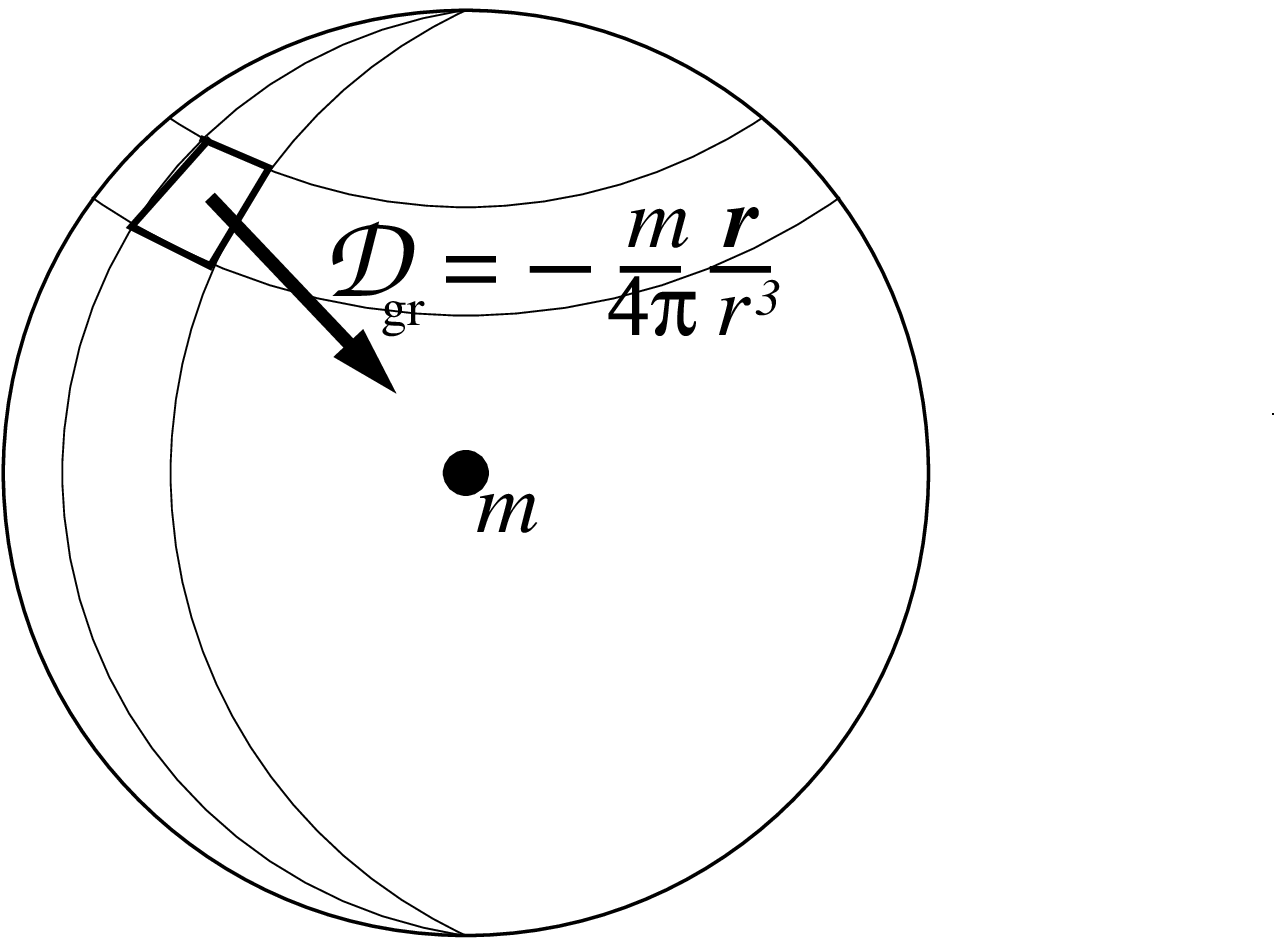}%{KottlerFig0_D.eps}
\caption{Gravitational excitation $\bm{\mathcal D}_{\text{gr}}$: We
  integrate $\bm{\partial}\cdot{\bm{\mathcal
      D}_{\text{gr}}}=-\rho_{\rm m}$ over a ball of radius $r$ and
  apply the Gauss theorem. For a point particle with mass $m$, this
  yields a gravitational excitation that is proportional to $m/r^2$.}
\end{figure}
\end{center}
$\bm{\mathcal D}_{\text{gr}}$ is the {\it active} part of the
gravitational field, created by the mass density scalar $\rho_{\rm
  m}$. Thus, the {\it conserved mass} (Lavoisier) is the source of
Newtonian gravity.

In electrodynamics, $\bm{\mathcal D}$ is a measurable quantity, since
we have positive and negative charges and $\bm{\mathcal D}$ can be
measured by the Maxwell double plates (because of the charge
separation). However, in gravity, we have only positive masses and the
gravitational excitation does not seem to have a direct operational
interpretation. Thus, $\bm{\mathcal D}_{\text{gr}}$ is of a more
formal nature than $\bm{\mathcal D}$. The fact that mass has only one
sign, in contrast to the two signs in the electric case, accounts for
this basic distinction between gravity and
electromagnetism.\footnote{If one requires the energy densities of the
  electromagnetic and the gravitational field to be positive, this
  entails that electromagnetism must be a vector and gravity a scalar
  or a tensor field of 2nd rank. Because of light deflection a pure
  scalar gravitational field is excluded, whereas a scalar admixture
  to the tensor field it is allowed for gravity (Jordan-Brans-Dicke
  theory). For these questions one should compare Deser \& Pirani
  \cite{Deser:1967zze,Deser:2004me}.}

%Tonti argues (priv.\ comm.) that this lack of an operational
%interpretation of $\bm{\mathcal D}_{\text{gr}}$ makes this quantity
%worthless. However, the vector potential $\bm{A}$ in magnetoscatics
%lacks such an operational definition, too. Still, however, it is an
%extremely useful concept (as underlined by Maxwell by choosing the
%first letter of the alphabet as its symbol).
%

%%%%%%%%%%%%%%%%%%%%%%%%%%%%%%%%%%%%%%%%%%%%%%%%%%%%%%%%%%%%%%%%%%%%%%%
The {\it passive} aspect of Newton's gravitational field can be
described as follows: In electrostatics, a force on a test charge $e$
is given by $\bm{F}=e\hspace{0.6pt}\bm{E}$ and the corresponding force
density by $ \bm{\mathfrak{F}}=\rho_{\rm el}\hspace{0.6pt}\bm{E}$.
This determines, in an operational way, the electric field strength
$\bm{E}$.  Similarly, the gravitational field strength $\bm{\Gamma}$
is defined via the force density
\begin{equation}\label{fieldstrength}
  \bm{\mathfrak{F}}_{\text{gr}} = \rho_{\rm m}\,\bm{\Gamma}\,,\qquad
  [\bm{\Gamma}] = {\frac  l {t^2}}= a = \text{acceleration}\,.
\end{equation}
According to Newton's attraction law, $\mathbf{\Gamma}$ is
conservative:
\begin{equation}\label{closedGamma}
\bm{\partial}\times\bm{\Gamma}=0\qquad {\rm or}\qquad
\bm{\Gamma}=-\,\bm{\partial}\phi.
\end{equation}
Following Kottler, we recognize that the equations (\ref{NGauss}) and 
(\ref{closedGamma}) are formulated independently of the metric of the 
Euclidean 3-space!

This premetric structure is even more clearly visible in exterior
calculus: for the 2-form ${\mathcal D}_{\text{gr}}$ of the
gravitational excitation, the {\it 3-form} $\varrho_{\rm m}$ of the
mass density, and the gravitational field strength 1-form $\Gamma$, we
have
\begin{equation}\label{NGauss1}
  d{\mathcal D}_{\text{gr}}=-\varrho_{\rm m}\,;\qquad
  {f}_{\text{gr}\,\a}=(e_\a\rfloor   \Gamma)\,
  \varrho_{\rm m}\,;\qquad d\Gamma=0\,,\qquad \Gamma=-d\phi\,.
\end{equation}
Here $e_\a$ is the local triad and $\rfloor$ denotes the interior
product of vector with an exterior form. Analogously, in vector
calculus, we find
\begin{equation}\label{NGauss2}
\bm{\partial}\cdot\bm{\mathcal D}_{\text{gr}} = -\,\rho_{\rm m}\,;\qquad  
 \bm{\mathfrak{F}}_{\text{gr}} = \rho_{\rm m}\bm{\Gamma}\,;\qquad
\bm{\partial}\times\,\bm{\Gamma}=0\,,\qquad \bm{\Gamma} = -\,\bm{\partial}\phi. 
\end{equation}
These are, according to Kottler, the premetric fundamental laws in
Newton's quasi-field theory of gravity.

%%%%%%%%%%%%%%%%%%%%%%%%%%%%%%%%%%%%%%%%%%%%%%%%%%%%%%%%%%%%%%%%%%%%%%%
The local and linear constitutive law relating the excitation
$\bm{\mathcal D}_{\text{gr}}$ to the field strength $\bm{\Gamma}$
requires a 3-dimensional Euclidean metric $g^{ab}$. Since
$\bm{\mathcal D}_{\text{gr}}$ is a vector density with components
${\mathcal D}_{\text{gr}}^{\,a}$ and $\bm{\Gamma}$ a covector with
components ${\Gamma}_b$, we find ($a,b=1,2,3$)
\begin{equation}\label{constitutive}
  {\mathcal D}_{\text{gr}}^{\,a}=\varepsilon_g\,\sqrt{-g}g^{ab}\,{\Gamma}_b
  \,,\qquad\varepsilon_g = {\frac 1{4\pi G}}\,,\qquad[\varepsilon_g] 
  = {\frac f{v^4}} = \frac{\text{force}}{\text{velocity}^4}\,.
\end{equation}
Here Newton's gravitational constant  enters, $G = 6.67\times
10^{-11}$m$^3/$kg\,s$^2$. The constitutive law
(\ref{constitutive}) $\bm{\mathcal D}_{\text{gr}} =
\varepsilon_g\,{\frak{g}}\,{\bm\Gamma}$ transforms Kottler's 3
premetric axioms in (\ref{NGauss2}) into a predictive physical theory,
namely the quasi-field theoretical version of Newton's gravitational
theory. Our results in (\ref{NGauss2}) and (\ref{constitutive}) are
consistent with the Tonti-diagram of the classical gravitational field
in \cite[p.401]{Tonti:2013}.

Substitution of the constitutive relation (\ref{constitutive}) into
the Coulomb-Gauss law (\ref{NGauss2}) yields the Poisson equation
\begin{equation}\label{Poisson1}
  \bm{\partial}\cdot\bm{\mathcal D}_{\text{gr}}  =
  \bm{\partial}(\varepsilon_g\,\cdot \frak{g}\,\bm{\Gamma}) =
  -\,\varepsilon_g\,\bm{\partial}\cdot\frak{g}\,  \bm{\partial}
  \,\phi = -\,\rho_{\rm m}\qquad\text{or}\qquad \varepsilon_g\,
  \Delta\phi=\rho_{\rm m}\,.
\end{equation}
Here $\Delta$ is the Laplace operator.  In exterior calculus, with the
3-dimensional metric-dependent Hodge star operator $^{\underline{\star}}$,
\begin{equation}\label{constitutive'}
{\mathcal D}_{\text{gr}} = \varepsilon_g{}\,^{\underline{\star}\,}\Gamma\,.
\end{equation}
\begin{equation}\label{Poisson1'}
  d\,{\mathcal D}_{\text{gr}}  = d(\varepsilon_g\,^{\underline{\star}}\Gamma)
  = - \,\varepsilon_g\,d\,^{\,\underline{\star}}d\,\phi  =-\varrho_{\rm m}\qquad
  \text{or}\qquad \varepsilon_g\,\Delta\phi=\,^{\underline{\star}}
  \varrho_{\rm m}\,,
\end{equation}
with the Laplacian for scalar fields $\Delta:=\,^{\underline{\star}}d
^{\,\underline{\star}}d\,$.  The Hodge star is the only metric
dependent quantity in these equations. For its dimension, see
footnote\footnote{If $\phi$ is a
  p-form and $[ds^2]= l ^2$, then in n dimensions
  $[^{\underline{\star}\,}\phi]=[\phi] l ^{n-2p}$. In particular,
  $[^{\underline{\star}\,}\rho] = [\rho]\, l ^{-3}$; this coincides with
  the dimension of the left-hand-side $\varepsilon_g\Delta\phi$,
  namely, $(f/v^4) l ^{-2}v^2 = m/ l ^3$, qed.}\!.

%%%%%%%%%%%%%%%%%%%%%%%%%%%%%%%%%%%%%%%%%%%%%%%%%%%%%%%%%%%%%%%%%%%%%
\section{Gravitomagnetism and critical assessment of Kottler's
  approach in gravity}\label{GEM}
%%%%%%%%%%%%%%%%%%%%%%%%%%%%%%%%%%%%%%%%%%%%%%%%%%%%%%%%%%%%%%%%%%%%%

In electrodynamics, there are two types of sources: electric charges
and electric currents (moving charges). Accordingly, there are two
types of phenom\-e\-na---the electric and magnetic ones. Their
existence and interrelation has a deep consequence: instead of the
clearly unphysical instantaneous action, electromagnetism is
propagating with a finite speed.

The natural idea that gravitation should propagate with a finite
speed, too, and that such a propagation process is substantiated by a
certain physical medium, is very old.  Already Laplace in his
``M\'echanique C\'eleste'' (1799) developed the model of a ``gravific
fluid,'' see the discussion of Whittaker \cite[Vol.\,1, pp.\
207--208]{Whittaker:1953}. In fact, even earlier, Newton himself
\cite{Newton:1692} wrote: ``That gravity should be innate inherent \&
\{essential\} to matter so that one body may act upon another at a
distance through a vacuum without the mediation of any thing else by
\& through which their action or force \{may\} be conveyed {}from one
to another is to me so great an absurdity that I believe no man who
has in philosophical matters any competent faculty of thinking can
ever fall into it. Gravity must be caused by an agent \{acting\}
consta\{ntl\}y according to certain laws, but whether this agent be
material or immaterial is a question I have left to the consideration
of my readers.'' One should also compare the fairly recent historical
investigation by Roseveare \cite{Roseveare:1982}.

Electrodynamics provided useful hints for the corresponding studies of
gravity.  Among the early attempts to modify Newton's gravity
accordingly, it is worthwhile to mention the investigations of
Holzm\"uller \cite{Holzmuller:1870} and Tisserand
\cite{Tisserand:1872,Tisserand:1890} who analyzed the possible
extension of Newton's gravity along the lines of Weber's
electrodynamics. Weber introduced velocity- and acceleration-dependent
corrections to the Coulomb force ${\frac
  {q_1q_2}{4\pi\varepsilon_0}}{\frac {\bm{r}}{r^3}} \left(1 - {\frac
    {\dot{r}^2}{2c^2}} + {\frac {r\,\ddot{r}}{c^2}}\right)$.  Doing
the same for Newton's gravitational force, it was possible to solve
the the problems of celestial mechanics and even to find the effect on
the precession of a planetary orbit (perihelion shift). However, for
the Mercury perihelion the result turned out to be incorrect, see
\cite[Vol.\,1, pp.\ 207--208]{Whittaker:1953}.

The problem was also addressed by Heaviside who asked himself
\cite{Heaviside:1893a}: ``Now what is there analogous to magnetic
force in the gravitational case? And if it has its analogue, what is
there to correspond with electric current?  At first glance it might
seem that the whole of the magnetic side of electromagnetism was
absent in the gravitational analogy.'' Taking Maxwell's
electrodynamics (and not Weber's one) as a guide, Heaviside came up
with the proposal to model the gravitational current as
\begin{equation}
  \bm{J}_{\rm m} = \rho_{\rm m}\bm{u}\,,\label{Ju}
\end{equation}
with the flow velocity $\bm{u}$ of the massive matter, and suggested
to relate this current (with a correction due to a ``displacement
current'' ) to the gravitomagnetic field $\bm{\mathcal H}_{\text{gr}}$
via the equation: ${\rm curl}\,\bm{\mathcal H}_{\text{gr}} =
-\,\bm{J}_{\rm m} + \dot{\bm{\mathcal D}}{}_{\text{gr}}$.

The beginning of the modern development of what is generally known as
the gravitoelectromagnetism goes back to the work of Thirring
\cite{Thirring:1918}, who noticed that the geodesic equation of a
massive test particle in general relativity can be recast into the
form of the equation of motion under the action of the Lorentz force
due to the gravitoelectric and gravitomagnetic fields. A modern and
easily accessible account can be found in Rindler
\cite[Sec.\,15.5]{Rindler:2001}, see particularly Rindler's table {}from
Eqs.(15.77) to (15.81).

In the weak field linear approximation Einstein's gravitational
equation can be formally written in a ``premetric'' form, compare
Table III,
\begin{eqnarray}\label{GEM1}
  \bm{\partial}\times\bm{\mathcal H}_{\text{gr}} - 
  \bm{\dot{\mathcal D}}{}_{\text{gr}} &=& 
  -\,\bm{J}_{\rm m}\,,\qquad \bm{\partial}\cdot\bm{\mathcal
    D}_{\text{gr}}   = -\,\rho_{\rm m}\,,\\
  \bm{\partial}\times\bm{E}_{\text{gr}} + \bm{\dot{B}}{}_{\text{gr}} &=& 0\,,
  \hspace{45pt} \bm{\partial}\cdot\bm{B}_{\text{gr}} = 0\,,\label{GEM2}
\end{eqnarray}
where the mass density $\rho_{\rm m}$ and the mass current
$\bm{J}_{\rm m}$ of (\ref{Ju}) are the sources of the gravitational
field. The gravitoelectric $\bm{E}_{\text{gr}} =
-\,\bm{\partial}\,\phi - \bm{\dot{\mathcal A}}$ and the
gravitomagnetic $\bm{B}_{\text{gr}} = \bm{\partial} \times\bm{\mathcal
  A}$ field strengths are linked to the gravitoelectric and
gravitomagnetic excitations $\bm{\mathcal D}_{\text{gr}}$ and
$\bm{\mathcal H}_{\text{gr}}$, respectively, by the constitutive
relations
\begin{equation}
  \bm{\mathcal D}_{\text{gr}} = \varepsilon_g\,\bm{E}_{\text{gr}}\,,\qquad 
  \bm{\mathcal H}_{\text{gr}} = {\frac 1{\mu_g}}\,\bm{B}_{\text{gr}}\,.
  \label{constGEM}
\end{equation}
The gravitolectric and the gravitomagnetic constants,
\begin{equation} 
  \varepsilon_g = {\frac 1{4\pi G}}\,,\qquad \mu_g 
  = {\frac {16\pi G}{c^2}}\,, 
\end{equation}
like the electric and magnetic constants in Maxwell's electrodynamics
(\ref{constM}), satisfy an analogous relation $\sqrt{\varepsilon_g\mu_g} =
2/c$.

When $\bm{\mathcal A} = 0$, we find $\bm{B}_{\text{gr}} = 0$ and the
system (\ref{GEM1})-(\ref{constGEM}) reduces to Kottler's premetric
formulation of Newton's gravity which we described in
Sec.\,\ref{Newton}, where we denoted $\bm{E}_{\text{gr}} =
\bm{\Gamma}$.

In order to make the scheme complete, we have to add the result of
Thirring \cite{Thirring:1918} for the force density on test matter:
\begin{equation}\label{forceGEM}
  \bm{\mathfrak{F}} = \rho_{\rm m}\bm{E}_{\text{gr}} + \bm{J}_{\rm
    m}\times  \bm{B}_{\text{gr}}\,.
\end{equation}

The complete analogy between electromagnetism, see Table III, and
gravitational theory, as represented in
  (\ref{GEM1}, \ref{GEM2}, \ref{constGEM}), is stunning. However, one
  has to be aware that in premetric electrodynamics the
electromagnetic fields $\bm{E}, \bm{B}$ and $\bm{\mathcal D},
\bm{\mathcal H}$ are pieces of the fundamental metric-free objects:
$F$ and $H$, respectively.

In general relativity (``Einstein gravity''), the
gravitolectromagnetic fields $\bm{E}_{\text{gr}}, \bm{B}_{\text{gr}}$
are constructed {}from the components of the spacetime metric
directly: $\phi = g_{00} - c^2,\, {\mathcal A}_a = -\,g_{0a}$. The
excitations $\bm{\mathcal D}_{\text{gr}}, \bm{\mathcal H}_{\text{gr}}$
are actually components of the Riemann curvature tensor, which is
found by differentiation of the connection (Christoffel symbols). The
Christoffels themselves are determined by differentiating the metric
$g$, namely ${\rm connection} \sim\,\partial\,(\,{\rm
  metric}\,)$. Thus, symbolically, gravitoelectromagnetism arises in
linearized Einstein gravity in the following way:
\begin{equation}
  \bm{E}_{\text{gr}},\bm{B}_{\text{gr}}\, \sim\,\partial\,(\,{\rm metric}\,),\quad
  \bm{\mathcal D}_{\text{gr}}, \bm{\mathcal H}_{\text{gr}}\,  \sim\,\partial
  \,(\,{\rm connection}\,)\sim\partial\partial\,(\,{\rm metric}\,)\,.
\end{equation}
In other words, in Einstein gravity the Riemannian metric is deeply
rooted in the gravitomagnetism, and the construction of the premetric
formulation of the gravity theory does not seem to be possible.

On the other hand, in the framework of gauge theories of gravity, it
is well-known that there is a so-called teleparallel equivalent of
general relativity, GR$_{||}$ (spoken ``GR teleparallel''), see
Blagojevi\'c et al. \cite[p.\,2, frontispiece]{Reader}, Itin
\cite{Itin:2001bp}, Obukhov \& Pereira \cite{Obukhov:2003}, and Maluf
\cite{Maluf:2013}. For ordinary matter, but excluding fermionic
fields, GR$_{||}$ is equivalent to Einstein gravity. However, the
gravitational potentials (belonging to the translations) are the
coframe (``tetrad'') and the connection becomes a teleleparallel
connection, both quantities that can be defined premetrically.  The
metric-dependent constitutive assumption would then be that the frame
is chosen {\it orthonormal}. Therefore, in the end the metric comes
in. However, the Kottler type framework with (\ref{GEM1}) and
(\ref{GEM2}) is uphold. The details of this GR$_{||}$ framework need
to be worked out in detail, but the general structures are well-known
and the program looks feasible to us. Accordingly, even in gravity, it
seems, Kottler was also on the right track.

%Of course, your eventual conclusions about Kottler and gravity are
%correct. We have premetric gravito-electromagnetic Maxwell equations, but
%
% (i) the grav. excitations are derived from the dual potential $\Gamma$
%     (Christoffels) and
%
%(ii) the grav. field strength are derived from the potential g, which
%     is the metric itself. Hence there is no way around the metric.
%
%Hence, my decisive question to you is: Can we absorb the factor 2 in the
%gravito-Lorentz force into the constitutive exations?

Back to Einstein gravity: the gravito-electromagnetic approach---for
a general over\-view, compare the textbooks
\cite{Schutz:2003nr,Ryder:2009}---can be extended also beyond the weak
field linear approximation
\cite{Jantzen:1992,Maartens:1998,Clark:2000,Maartens:2008,Boos:2014}. It
plays an important role in modern gravitational theory, in particular
with respect to the discussion of the inertial drag and spin effects
\cite{Ruggiero:2002,Mashhoon:2007}.

%%%%%%%%%%%%%%%%%%%%%%%%%%%%%%%%%%%%%%%%%%%%%%%%%%%%%%%%%%%%%%%%%%%%%
\section{Linear and local electrodynamics}\label{Linear}
%%%%%%%%%%%%%%%%%%%%%%%%%%%%%%%%%%%%%%%%%%%%%%%%%%%%%%%%%%%%%%%%%%%%%

After this somewhat complicated case with gravity, let us come back to
electrodynamics. In Sec.\,\ref{constants} we saw, see (\ref{const&}),
$H_{ij}=\frac{1}{2}\kappa_{ij}{}^{kl}F_{kl}$, that, on the premetric
level, we can assume a medium with local and linear response in order
to complete the fundamental equations of electrodynamics to a fully
predictive theory: in components, it comprises the Maxwell equations
(\ref{bothMax}) and the local and linear law (\ref{const&}).

Our conclusion contradicts, however, the verdict of Tonti
\cite{Tonti:2016}. Whereas he agrees that the Maxwell equations can be
formulated premetrically, he insists that ``all constitutive equations
require the metric.'' No doubt, he is mostly correct with this
stipulation. The collection of constitutive laws assembled in Tonti's
book \cite{Tonti:2013} seems to attest to his belief. 

There are interesting exceptions known, though: (i) The
phenomenological constitutive law that describes the quantum Hall
effect within (1+2)-dimensional electrodynamics, see Bieri \&
Fr\"ohlich \cite[Eq.12]{Bieri:2010za} and Hehl \& Obukhov
\cite[Eq.(B.4.60]{Birkbook}. This should not come as a surprise when
in condensed matter circles the notion of the {\it topological
  superconductors} is prevalent. (ii) The axionic part of the
constitutive law of the multiferroic chromium sesquioxide Cr$_2$O$_3$
is premetric, see \cite[Eq.(17)]{Hehl:2007ut}. Accordingly, these laws
are topological and independent of the metric. Other examples are
known.  In any case, Tonti's belief that all constitutive law require
a metric is not universally valid. This will be also demonstrated in
Sec.\,\ref{Metric} where a metric is {\it derived} {}from a premetric law
by requiring certain supplementary postulates.

%%%%%%%%%%%%%%%%%%%%%%%%%%%%%%%%%%%%%%%%%%%%%%%%%%%%%%%%
\subsection{A look at the history...}
%%%%%%%%%%%%%%%%%%%%%%%%%%%%%%%%%%%%%%%%%%%%%%%%%%%%%%%%

...of our subject may widen our view. Already in 1910, Bateman
\cite{Bateman:1910} investigated local and linear materials. In his
own notation, his constitutive equations were
\cite[Eq.(II)]{Bateman:1910}
\begin{eqnarray}\label{constBateman}
  -B_1&=&\kappa_{11}H_1+\kappa_{12}H_2+\kappa_{13}H_3+\kappa_{14}D_1
  +\kappa_{15}D_2+\kappa_{16}D_3\,,\nonumber\\
  &&\hspace{-5pt} ...\hspace{28pt}...\hspace{35pt}...\hspace{35pt}...
  \hspace{35pt}...\hspace{35pt}...\nonumber\\
  E_3&=&\kappa_{61}H_1+\kappa_{62}H_2+\kappa_{63}H_3+\kappa_{64}D_1
  +\kappa_{65}D_2+\kappa_{66}D_3\,,
\end{eqnarray}
with the symmetry requirement $\kappa_{rs}=\kappa_{sr}$, $r,s=1,\dots
6$, which leaves only 21 independent components. Besides the
permeabilities (here $\kappa_{11},\kappa_{12},\kappa_{13}$) and the
permittivities (here $\kappa_{64},\kappa_{65},\kappa_{66}$), also
magneto-electric cross terms (the rest of the terms) were assumed, the
latter of which were not discovered before 1961 by Astrov
\cite{Astrov} (here $\kappa_{61},\kappa_{62},\kappa_{63}$) )and by
Rado \& Folen \cite{RadoFolen} (here
$\kappa_{14},\kappa_{15},\kappa_{16}$), see also O'Dell
\cite{O'Dell:1970}.

Nowhere did Bateman mention the metric explicitly. Also his Maxwell
(vacuum) equations had a premetric form. Thus, he already anticipated
Kottler's program to some extend. However, he did not think that his
assumptions were related to physics \cite{Bateman:1910}: {\it ``These
  conditions} [the constitutive relations (\ref{constBateman})] {\it
  may not correspond to anything occurring in nature; nevertheless
  their investigation was thought to be of some mathematical interest
  on account of the connection which is established between
  line-geometry and the theory of partial differential equations.}''
Then, {\it ``The general Kummer's surface appears to be the wave
  surface for a medium of a purely ideal character...''}  To this last
remark, we will come back in the next section, see Baekler et
al.\ \cite{Baekler:2014kha}.\bigskip

In 1925, Tamm \cite{Tamm:1925} used modern 4-dimensional tensor
analysis, and he assumed the local and linear law
$f^{ij}=s^{ijpq}F_{pq}$, which is equivalent to (\ref{const&}). In our
notation, it reads
\begin{equation}\label{Hcheck}
  \check{\mathcal H}^{ij}=\frac 12\chi^{ijkl}F_{kl}\,,
\end{equation}
with the excitation tensor density $\check{\mathcal
  H}^{ij}:=\frac 12\epsilon^{ijkl}H_{kl}$ and the electromagnetic moduli
$\chi^{ijkl}:=\frac 12\epsilon^{ijmn}\kappa_{mn}{}^{kl}$. In his ansatz,
Tamm, like Bateman, had 21 independent components. In his actual
applications, he used, however, only 4 of them. This state of the art
was summarized by the monographs of Post \cite{Post:1962} (see also
Post \cite{Post:1979}) and O'Dell \cite{O'Dell:1970}, with the only
proviso that Post killed, by mistake, one of the 21 independent
components of $\chi^{ijkl}$ (``Post constraint'').

The symmetry of Bateman's $\kappa_{rs}$ or, equivalently, the symmetry
$\chi^{ijkl}=\chi^{klij}$ was {\it dropped} for the first time by
Serdyukov, Semchenko, Tretyakov, and Sihvola \cite{Serdyukov:2001}, as
far as we know. Thus, they arrived at the 36 independent components
postulated in (\ref{const&}). In this way dissipative processes can be
included. A systematic study of the 36 components of $\chi^{ijkl}$
was performed by two of us \cite{Birkbook}. In particular, an
irreducible decomposition was achieved for $\chi^{ijkl}$ for the first
time. It was cut into a principal part (20 components), a skew part
(15 independent components), and an axion part (1 independent
component). Incidentally, the Maxwell-Lorentz vacuum law in contained
in the principal part in the form of
\begin{equation}\label{vacuumChi}
  \chi^{ijkl}=2\lambda_0\,\sqrt{-g}g^{i[k}g^{l]j}\,,
\end{equation}
see (\ref{vacuum}). Subsequently, $\chi^{ijkl}$ with 36 independent
moduli were investigated, amongst others, by Lindell
\cite{Lindell:2004,Lindell:2005,Lindell:2015}, Favaro \cite{PhD}, Ni
\cite{Ni:2015rdf}, and Pfeifer and Siemssen\cite{Pfeifer:2016har}.

If one has moving bodies around, also the 4-velocity of the body could
be included into such considerations, as was early observed by
Schmutzer \cite{Schmutzer:1968}, see in this context also Balakin et
al. \cite{Balakin:2015kja, Balakin:2016cbe}.

\subsection{Local and linear response}\label{Llr}

After this look back in the history of the constitutive law, let us
come back to (\ref{const&}). It reads
\begin{equation}\label{const&*}
  {H}_{ij} = \frac 12\kappa_{ij}{}^{kl}F_{kl}\,.
\end{equation}
It can be recast into the decomposed form
\begin{eqnarray}\label{const1}
  {\mathcal H}_a &=& -\,{\mathfrak C}^{b}{}_a\,E_b + {\mathfrak B}_{ba}\,B^b\,,\\ 
 {\mathcal D}^a &=& -\,{\mathfrak A}^{ba}\,E_b
 + {\mathfrak D}_{b}{}^a\,B^b\,.\label{const2}
\end{eqnarray}
The generalized permittivity matrix ${\mathfrak A}^{ba}$ and the generalized 
{\it im\/}permeability matrix ${\mathfrak B}_{ba}$ and the magneto-electric 
matrices ${\mathfrak C}^{b}{}_a$ and ${\mathfrak D}_{b}{}^a$ are constructed 
{}from the components of the constitutive tensor,
\begin{eqnarray}
{\mathfrak A}^{ba} = {\frac 12}\epsilon^{bcd}\kappa_{cd}{}^{0a},\qquad
{\mathfrak B}_{ba} = {\frac 12}\epsilon_{bcd}\kappa_{0a}{}^{cd},\label{ABconst}\\
{\mathfrak C}^{b}{}_a = \kappa_{0a}{}^{0b},\qquad {\mathfrak D}_{b}{}^a = 
{\frac 14}\epsilon^{acd}\epsilon_{bef}\kappa_{cd}{}^{ef}.\label{CDconst}
\end{eqnarray}

The twisted constitutive tensor $\kappa_{ij}{}^{kl}(x)$ of type
$\left[^{2}_{2}\right]$ has 36 independent components. We can
decompose this object into several irreducible parts. Define
\begin{equation}
\kappa_i{}^k := \kappa_{il}{}^{kl}\,,
\end{equation}
with 16 independent components. 
The second contraction yields the twisted scalar function
\begin{equation}\label{trace}
\kappa := \kappa_k{}^k = \kappa_{kl}{}^{kl}\,
\end{equation}
(also called pseudo- or axial-scalar).  The traceless piece
\begin{equation}\label{tracefree}
\not\!\kappa_i{}^k := \kappa_i{}^k - {\frac 1 4}\,\kappa\,\delta_i^k
\end{equation}
has 15 independent components. Three irreducible pieces then read:
\begin{eqnarray}
  \kappa_{ij}{}^{kl} &=& {}^{(1)}\kappa_{ij}{}^{kl} +\nonumber
  {}^{(2)}\kappa_{ij}{}^{kl} + {}^{(3)}\kappa_{ij}{}^{kl} \\ &=&
  {}^{(1)}\kappa_{ij}{}^{kl} +
  2\!\not\!\kappa_{[i}{}^{[k}\,\delta_{j]}^{l]} + {\frac 1
    6}\,\kappa\,\delta_{[i}^k\delta_{j]}^l\,.\label{kap-dec}
\end{eqnarray}
Besides the {\it principal part} ${}^{(1)}\kappa_{ij}{}^{kl}$, we have
the {\it skewon} and the {\it axion} fields
\begin{equation}
\!\not\!S_i{}^j := -\,{\frac 1 2}\!\not\!\kappa_i{}^j,\qquad
\alpha := {\frac 1 {12}}\,\kappa.\label{Salpha}
\end{equation}
Accordingly, the general local and linear constitutive relation
(\ref{const&}) reads explicitly
\begin{equation}\label{crypto2a}
  H_{ij}=\frac 12\,^{(1)}\kappa_{ij}{}^{kl}\,F_{kl}+2\, {\!\not
    \!S}_{[i}{}^kF_{j]k}+\alpha\,F_{ij}\,.
\end{equation} 
Thus, we split $\kappa_{ij}{}^{kl}$ according to $36 = 20 + 15 + 1$.

Let us look into the properties of the three irreducible
pieces. First, we study the principal part. By construction,
${}^{(1)}\kappa_{ij}{}^{kl}$ is totally traceless:
\begin{equation}
{}^{(1)}\kappa_{il}{}^{kl} = 0.\label{notrace}
\end{equation}
It is more nontrivial to verify the double-duality property
\begin{equation}
{}^{(1)}\kappa_{ij}{}^{kl} = {\frac 14}\epsilon_{ijpq}\epsilon^{klmn}\,
{}^{(1)}\kappa_{mn}{}^{pq},\label{double}
\end{equation}
but in fact this is a direct consequence of (\ref{notrace}). Following
(\ref{ABconst}) and (\ref{CDconst}), we introduce the 3-dimensional
blocks of the principal part
\begin{eqnarray}\label{epsmu}
  \varepsilon^{ba} &=& -\,{\frac 12}\epsilon^{bcd}{}^{(1)}\kappa_{cd}{}^{0a},\qquad
  \mu^{-1}_{ba} = {\frac 12}\epsilon_{bcd}{}^{(1)}\kappa_{0a}{}^{cd},\\
  \gamma^{b}{}_a &=& {}^{(1)}\kappa_{0a}{}^{0b} = {\frac
    14}\epsilon^{bcd}\epsilon_{aef}\, {}^{(1)}\kappa_{cd}{}^{ef}.\label{gab}
\end{eqnarray} 
Taking into account (\ref{notrace}) and (\ref{double}), we can then
straightforwardly demonstrate the symmetry properties
\begin{equation}\label{sym}
  \varepsilon^{ab}=\varepsilon^{ba},\qquad
  \mu^{-1}_{ab}=\mu^{-1}_{ba},\qquad \gamma^c{}_c=0. 
\end{equation}
Thus we indeed verify that the total number of independent components
($\varepsilon^{ab}$, $\mu^{-1}_{ab}$, $\gamma^a{}_b$) of the principal
part ${}^{(1)}\kappa_{ij}{}^{kl}$ is $6 + 6 + 8 = 20$.

Turning to ${}^{(2)}\kappa_{ij}{}^{kl}$, it is convenient to
parametrize the skewon part by
\begin{equation}\label{prrameterS}
  \!\not\!S_i{}^j= \left(\begin{array}{cc}-s_c{}^c & m^a \\ n_b &
      s_b{}^a \end{array}\right)\,.
\end{equation}
The total number of parameters ($m^a$, $n_b$, $s_b{}^a$) correctly yields 
$3 + 3 + 9 = 15$ skewon degrees of freedom. 

We thus eventually reveal the fine structure of the constitutive tensor
\begin{eqnarray}
 \kappa = \!\underbrace{\left(\begin{array}{cc}
        \gamma^b{}_a & \mu_{ab}^{-1} \\ -\varepsilon^{ab} &
        \gamma^a{}_b \end{array}\right)}_{ principal\>{\rm part\> 20\>
      comp.}}\! + \underbrace{ \left(\begin{array}{cc} - s_a{}^b
        +\delta_a^b s_c{}^c &\>\; - \hat{\epsilon}_{abc}m^c \\ 
        \epsilon^{abc}n_c & s_b{}^a - \delta_b^a s_c{}^c
  \end{array}\right)}_{ skewon\> {\rm part\> 15\> comp.}}\!
+ \underbrace{\alpha\left(\begin{array}{cc}\delta_a^b&0\\
 0&\delta_b^a\end{array}\right)}_{ axion\>{\rm part\> 1 \>comp.}}.
\label{CR''''}
\end{eqnarray}
Accordingly, (\ref{const1}) and (\ref{const2}) read explicitly as follows:
\begin{eqnarray}\label{explicit'}
  {\mathcal H}_a\!&=\!&\left( \mu_{ab}^{-1} - \hat{\epsilon}_{abc}m^c
  \right) {B}^b\, - \alpha\,E_a %\nonumber\\&&\qquad\qquad  
+\left(- \gamma^b{}_a + s_a{}^b - \delta_a^b
    s_c{}^c\right)E_b \,,\\ {\mathcal D}^a\!&=\!&\left(
    \varepsilon^{ab}\hspace{4pt}  - \, \epsilon^{abc}\,n_c\right)E_b\,
 + \alpha\,B^a %\nonumber\\&&\qquad\qquad 
+\left(\hspace{9pt} \gamma^a{}_b 
  + s_b{}^a - \delta_b^a s_c{}^c\right) {B}^b \,.\label{explicit''}
\end{eqnarray}
Recall that $\varepsilon^{ab}=\varepsilon^{ba}$,
$\mu^{-1}_{ab}=\mu^{-1}_{ba}$, and $\gamma^c{}_c=0$. Incidentally,
$\alpha$ is a 4-dimensional (axial) scalar, whereas $s_c{}^c$
(summation over $c$) is only a 3-dimensional scalar.

The physical meaning of the various pieces of the general linear
constitutive law is clear: $\varepsilon^{ab}$ stands for a generalized
permittivity, $\mu^{-1}_{ab}$ encodes the impermeability properties,
and the cross-term $\gamma^a{}_b$ accounts for the magneto-electric
moduli and is related to the Fresnel-Fizeau effects.  The skewon
contributions $m^c,n_c$ are responsible for electric and magnetic
Faraday effects, respectively, whereas the skewon terms $s_a{}^b$
describe optical activity. We underline that nowhere in our
considerations a metric was used. This analysis is completely
premetric. 

The premetric approach of electrodynamics has also been developed over
the years by Delphenich
\cite{Delphenich:2006bm,Delphenich:2009,Delphenich:2015} and, more
recently, by Cabral \& Lobo \cite{Cabral:2016yxh}. Rumpf
\cite{Rumpf:2015} and Zhang et al.\cite{Zhang:2016}, for instance,
discuss realistic engineering applications for
(\ref{explicit'},\ref{explicit''}), but without using the the skewon
part. Similar constitutive laws are applied by de Carvalho
\cite{deCarvalho:2015toa} to the relativistic electron gas.

%%%%%%%%%%%%%%%%%%%%%%%%%%%%%%%%%%%%%%%%%%%%%%%%%%%%%%%%%%%%%%%%%%%%%
\subsection{Axion part of the magnetoelectric moduli found in the
  multiferroic Cr$_2$O$_3$}\label{Axion}
%%%%%%%%%%%%%%%%%%%%%%%%%%%%%%%%%%%%%%%%%%%%%%%%%%%%%%%%%%%%%%%%%%%%%

The decomposition of the constitutive law, as manifest in
(\ref{explicit'}) and (\ref{explicit''}), immediately suggests to look
for experimental evidence of the 4d pseudoscalar $\alpha$, the axion
part of the constitutive tensor, see
\cite{Hehl:2004tk,Obukhov:2005kh}. It is a specific magnetoelectic
piece, that is, after the discovery of magneto-electricity in 1961, it
was within the reach of experimentalists. Two of us joint forces with
two experimental colleagues
\cite{Hehl:2007ut,Hehl:2007jy,Hehl:2009eqa}, who had the corresponding
measurements already in their drawer. It was the multiferroic chromium
sesquioxide Cr$_2$O$_3$ that carries a small axion piece of at most
$\alpha\sim 10^{-4}\lambda_0$ at around 275\hspace{1pt}K. Thus, in
2008, the Post constraint was disproved, that is, an axion part, even
if rare and small, can exist.

Model calculations for the axion piece were made by Essin, Moore, and
Vanderbilt \cite{Essin:2008rq}. New experiments with Cr$_2$O$_3$ were
described, for example, by Seki et al.\ \cite{Seki:2015} and wave
propagation studied by Lindell, Sihvola, and Favaro
\cite{Lindell:2016a}.

No other material than Cr$_2$O$_3$ has been described in the meantime
with a nonvanishing $\alpha$. Rosch \cite{Rosch:2016} has suggested to
one of us that the copper selenide Cu$_2$OSeO$_3$, an insulator, due
its symmetries, could possibly carry an $\alpha\ne 0$. Linear optical
properties of this chiral magnet Cu$_2$OSeO$_3$ were recently
investigated by Versteeg et al.\ \cite{Versteeg}. Because of the
chiral properties of this material, the existence of a nonvanishing
axion piece would be plausible.

If vacuum electrodynamics is amended by an axion piece, one speaks of
{\it axion electrodynamics.} This theory goes back to ideas that Ni
developed in the 1970s \cite{Ni73,Ni77}. Corresponding applications
were discussed recently by Balakin \cite{Balakin:2016ygw}, for
example. Axion electrodynamics can also be used to discuss the
response of materials to photons at low energy. Optical effects, like
the Faraday and the Kerr rotation can be calculated with its help, see
Wu et al.\ \cite{Wu:2016oxw}.

As we described in \cite{Hehl:2007ut}, the axion piece $\alpha$ is
akin to a Tellegen material. Some new insight in this connection is
provided by Proden\^cio et al.\ \cite{Prudencio:2014}.

%%%%%%%%%%%%%%%%%%%%%%%%%%%%%%%%%%%%%%%%%%%%%%%%%%%%%%%%%%%%%%%%%%%%%
\subsection{Premetric reciprocity analysis}\label{Reciprocity}
%%%%%%%%%%%%%%%%%%%%%%%%%%%%%%%%%%%%%%%%%%%%%%%%%%%%%%%%%%%%%%%%%%%%%

In order to demonstrate further the power of our approach, we will show that
already on this premetric level the notion of reciprocity can be
applied. We recall that, in a very general sense, the reciprocity of
the electromagnetic media means that after changing the source of the
field with the measuring device, the reading of the measuring device
remains the same.

Let us consider a premetric generalization of the ``reaction" quantity
\cite{Rumsey}: 
\begin{equation}\label{reaction}
\int\limits_V ({}^{(1)}\!E\wedge{}^{(2)}\!J - {}^{(2)}\!E\wedge{}^{(1)}\!J).
\end{equation}
Here the integral is taken over the spatial volume $V$, and the 1-form of
the electric field ${}^{(1)}\!E = {}^{(1)}\!E_a dx^a$ is created by the source 
2-form of the electric current ${}^{(1)}\!J = {}^{(1)}\!J^a\,\epsilon_a$, 
whereas the field ${}^{(2)}\!E$ is created by the current ${}^{(2)}\!J$.

The concept of reaction in the electromagnetic theory is used for the analysis
of electromagnetic wave propagation in media and it has important applications
for the formulation of the boundary value problem and computation of the 
scattering and transmission coefficients. 

Using the premetric Maxwell equations, see Table III, we derive
\begin{equation}
{}^{(1)}\!E\wedge{}^{(2)}\!J = -\,d({}^{(1)}\!E\wedge{}^{(2)}\!{\cal H})
- {}^{(2)}\!{\cal H}\wedge{}^{(1)}\!\dot{B} - {}^{(1)}\!E\wedge{}^{(2)}
\!\dot{\cal D}. 
\end{equation}

Making use of the constitutive relations (\ref{const1}) and (\ref{const2}), we find
\begin{eqnarray}\label{reciprocity}
&\int\limits_V ({}^{(1)}\!E\wedge{}^{(2)}\!J - {}^{(2)}\!E\wedge{}^{(1)}\!J)
%= \nonumber\\ &&
= \int\limits_{\partial V} ({}^{(2)}\!E\wedge{}^{(1)}\!{\cal H} - {}^{(1)}
\!E\wedge{}^{(2)}\!{\cal H})& \nonumber\\
& + \int\limits_V \left[2{\mathfrak A}^{[ba]}{}^{(1)}\!E_a{}^{(2)}\!\dot{E}_b
- 2{\mathfrak B}_{[ba]}{}^{(1)}\!B^a{}^{(2)}\!\dot{B}^b \right.& \nonumber\\ 
&\left. + \,({\mathfrak C}^{b}{}_a + {\mathfrak D}_{a}{}^b)({}^{(2)}\!E_b
{}^{(1)}\!\dot{B}^a - {}^{(1)}\!E_b{}^{(2)}\!\dot{B}^a)\right]\epsilon.& 
\end{eqnarray}

The reciprocity requires that the ``reaction'' (\ref{reaction}) vanishes
for all sources. Usually, $V$ is the whole space, and the surface 
integral over the boundary $\partial V$ disappears by assumption. 
As a result, we find that the reciprocal media are characterized 
by the following properties of the constitutive matrices:
\begin{equation}
{\mathfrak A}^{ab} = {\mathfrak A}^{ba},\qquad
{\mathfrak B}_{ab} = {\mathfrak B}_{ba},\qquad
{\mathfrak C}^{b}{}_a + {\mathfrak D}_{a}{}^b = 0.\label{recABCD}
\end{equation}

Recalling the irreducible decomposition of the constitutive tensor,
we thus see that neither of the three irreducible parts is totally
reciprocal. In the principal part, the magneto-electric susceptibilities
$\gamma^a{}_b$ are nonreciprocal, whereas in the skewon part the 
Faraday-responsible parameters $m^a, n_a$ are nonreciprocal. The 
axion, being a special (4-dimensional isotropic) case of magneto-electricity,
is also nonreciprocal. Schematically, we have 
\begin{equation}
{}^{(1)}\kappa = \left(\begin{array}{cc}n&r\\ r&n\end{array}\right),\quad
{}^{(2)}\kappa = \left(\begin{array}{cc}r&n\\ n&r\end{array}\right),\quad
{}^{(3)}\kappa = \left(\begin{array}{cc}n&0\\ 0&n\end{array}\right).
\end{equation}
Here ``$r$'' and ``$n$'' stands for reciprocal and non-reciprocal, respectively.

In general, a reciprocal medium is characterized by the electric and
magnetic permeabilities $\varepsilon, \mu$ and by the skewonic
susceptibilities $s^a{}_b$.  When the former are purely real and the
latter is purely imaginary, the matter is non-dissipative. It is
sometimes called a chiral bi-anisotropic medium.  Imaginary parts of
permittivity and permeability and the real part of the skewon are
responsible for dissipation. It was demonstrated for complex media
(see \cite{Mcisaac,Kam1}, e.g.) that reciprocity is related, in the
broad sense, to the Onsager-Casimir time-reversal invariance
\cite{Onsager,Casimir} of the equations of motion of a physical
system. However the complete understanding of the details of this
relation is still absent, cf.\ \cite{Rado:1973,Tretyakov:2002} and
\cite{LakhtakiaDepine}.  Nonreciprocity of magneto-electric materials
is featured, in particular, in \cite{Krowne1,Krowne2}. For a general
discussion of the reciprocity in bi-anisotropic media one can refer to
the books \cite{Altman:2011,Lindell:2004,Lindell:2015}, for
applications see also Monzon \cite{Monzon1,Monzon2}. \medskip

%%%%%%%%%%%%%%%%%%%%%%%%%%%%%%%%%%%%%%%%%%%%%%%%%%%%%%%%%%%%%%%%%%%%%
\section{Waves, rays, generalized Fresnel equation}\label{Waves}
%%%%%%%%%%%%%%%%%%%%%%%%%%%%%%%%%%%%%%%%%%%%%%%%%%%%%%%%%%%%%%%%%%%%%

Having set up the premetric framework with the Maxwell equations and
the 36 independent components of the electromagnetic response tensor,
we can now draw the consequences of this layout. We take here the
version (\ref{Hcheck}) for the response law. The immediate reaction of
most physicists would be to look for the wave phenomena related to the
theory under investigation and---turning to the geometrical optics
limit---to study the emerging light rays. And this is exactly what we
will do; for a classical text, see Born \& Wolf
\cite{Born:2001}.\medskip

%%%%%%%%%%%%%%%%%%%%%%%%%%%%%%%%%%%%%%%%%%%%%%%%%%%%%%%%%%%%%%%%%
\subsection{Hadamard's method, Tamm-Rubilar tensor density}
%%%%%%%%%%%%%%%%%%%%%%%%%%%%%%%%%%%%%%%%%%%%%%%%%%%%%%%%%%%%%%%%%

We implemented this idea by using a standard method due to {Hadamard}
\cite{Hadamard} by considering a wave surface with a continuous
electromagnetic field strength $F$, but the derivative of $F$ has a
jump. The direction of the jump is given by the wave covector
$q_i$. By integration, we can create thereby the wave vectors as
tangents to the rays \cite{Birkbook}.

This was done roughly since the year 2000 and the results have been
reviewed in the book of two of us \cite{Birkbook}, see also Obukhov et
al.\ \cite{Obukhov:2000nw}, Hehl et al.\ \cite{Hehl:2002hr}, Rubilar
\cite{Rubilar:2007qm} (and for nonlinear electrodynamics Obukhov et
al.\ \cite{Obukhov:2002xa}). The outcome is a generalized Fresnel
equation, see \cite[Eq.(D.2.23)]{Birkbook}, which is (algebraically)
quartic in the wave covector $q_i$:
\begin{eqnarray}\label{Fresnel}
  {\cal G}^{ijkl}[\chi]\,q_iq_jq_kq_l =0\,.
\end{eqnarray}
\begin{figure}\label{Fig1}
\includegraphics[width=12cm]{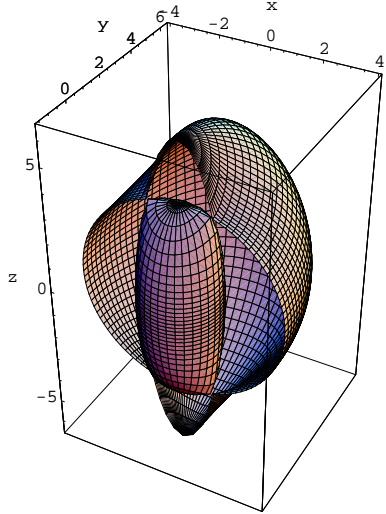}%{KottlerFig1_aniW.eps}
\caption{This is Fresnel surface for the anisotropic medium without
  skewon part. It is a Kummer surface, see Obukhov et al.\
  \cite{Obukhov:2004zz}.}
\end{figure}
\begin{figure}\label{Fig2}
\includegraphics[width=12cm]{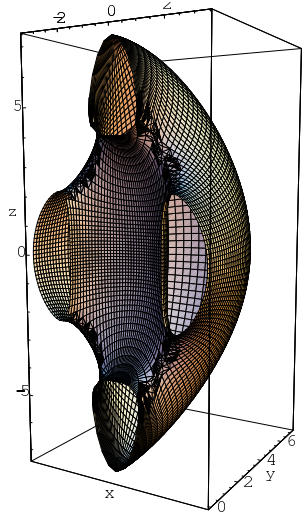}%{KottlerFig2_holes.eps}
\caption{Here the skewon part is present: $^{(2)\!}\chi^{ijkl}\ne
  0$. The two branches merge into one, and the points where the old
  branches in Fig.\,2 touched are replaced by the holes---the wave
  vectors do not go though the holes. This illustrates the typical
  effect of a skewon: damping of wave propagation in certain
  directions (or everywhere), see Obukhov et al.\
  \cite{Obukhov:2004zz}.}
\end{figure}
This equation is governed by the {\it Tamm-Rubilar} tensor density,
which is cubic in $\chi$, see
\cite[Eq.(D.2.22)]{Birkbook}:\footnote{We use here the left diamond
  dual ${^\diamond}\chi_{ij}{}^{kl}:=\frac 12 \epsilon_{ijab}
  \chi^{abkl}$ and the right diamond dual $ \chi^{\diamond\,
    ij}{}_{kl}:=\frac 12 \chi^{ijcd}\epsilon_{cdkl}$ as convenient
  abbreviations.}
\begin{eqnarray}\label{TammRubilar}
  {\cal G}^{ijkl}[\chi]:=\frac{1}{3!}\,^{\diamond}\chi_{ab}{}^{c(i\,}\chi^{j|ad|k}
\chi^{\diamond\,l)b}{}_{cd}\,.
\end{eqnarray}
Note that ${\cal G}^{ijkl}$ is totally symmetric ${\cal
  G}^{ijkl}={\cal G}^{(ijkl)}$; thus, it has 35 independent
components. Moreover, the axion part
$^{(3)\!}\chi^{ijkl}=\chi^{[ijkl]}$ drops out of (\ref{TammRubilar})
and, accordingly, also of (\ref{Fresnel}), that is, the axion part
cannot be seen in the geometric optics limit. Curiously, ${\cal
  G}^{ijkl}={\cal G}^{(ijkl)}$, which is cubic in $\chi^{ijkl}$, as
well as $(\chi^{ijkl}-\,{^{(3)\!}\chi^{ijkl}})$, have the same number of
independent components, namely 35---this is probably not by chance.

The generalized Fresnel equation (\ref{Fresnel}) spans the Fresnel
surface. We present three typical cases for visualizing it, see also
Favaro \cite{Favaro:2016pej}. In Fig.\,4 a normal Fresnel surface for
an anisotropic crystal, in Fig.\,5 the presence of a hypothetical
skewon part is allowed, and in Fig.\,6 the extreme case of a
hypothetical metamaterial with moduli \cite{Favaro:2015jxa}, namely
%\begin{align} \label{eq:noinf1}
%  \varepsilon^{ab} &= - {\frac{1}{4}}\left(\begin{array}
%  {ccc}1+\sqrt{3} & 0 & 0\\  0& 1+\sqrt{3} & 0\\ 0& 0& 4-2\sqrt{3}
%  \end{array}\right)  ,\\
%\mu^{-1}_{ab} &=\hspace{9pt} {\frac{1}{4}}\left(\begin{array}
%    {ccc}1+\sqrt{3} & 0 & 0\\ 0& 1+\sqrt{3} & 0\\ 0& 0& 4-2\sqrt{3}
%  \end{array}\right), \label{eq:noinf2}\\ \label{eq:noinf3}
%  \gamma^{a}{}_{b} &=
%\frac{\sqrt{3}}{4} {\left( {1+\sqrt{3}}\right)}\left(\begin{array}{ccc}
%  1 & 0 & 0\\ 0 & - 1 & 0\\ 0 & 0& 0 \end{array}\right).
%\end{align}
%Alternative presentation of the moduli for Fig.\,3:
\begin{align} \label{eq:noinf1}
  \varepsilon^{ab} &= \hspace{9pt}\frac 14(1-\sqrt{3})\left(\begin{array}
    {ccc}1&0&0\\ 0&1&0\\0&0&-2 \end{array}\right)+\frac 12\left(\begin{array}
    {ccc}1&0&0\\ 0&1&0\\0&0&1\end{array}\right)\,,\\
\mu^{-1}_{ab} &=- \frac 14(1-\sqrt{3})\left(\begin{array}
    {ccc}1&0&0\\ 0&1&0\\0&0&-2 \end{array}\right)-\frac 12\left(\begin{array}
    {ccc}1&0&0\\ 0&1&0\\0&0&1\end{array}\right)\,, \label{eq:noinf2}\\ \label{eq:noinf3}
  \gamma^{a}{}_{b} &=
\frac{\sqrt{3}}{4} {( {1+\sqrt{3}})}\left(\begin{array}{ccc}
  1 & 0 & 0\\ 0 & - 1 & 0\\ 0 & 0& 0 \end{array}\right).
\end{align}

This yields the maximal possible number of 16 real singular points of
a Kummer surface. Note that in this material the skewon part vanishes
by assumption. It seems that the Kummer surfaces are destroyed by the
skewon part, see Fig.\,4, but we have no general proof for that.
\begin{figure}\label{Fig3}
\includegraphics[width=0.82\columnwidth]{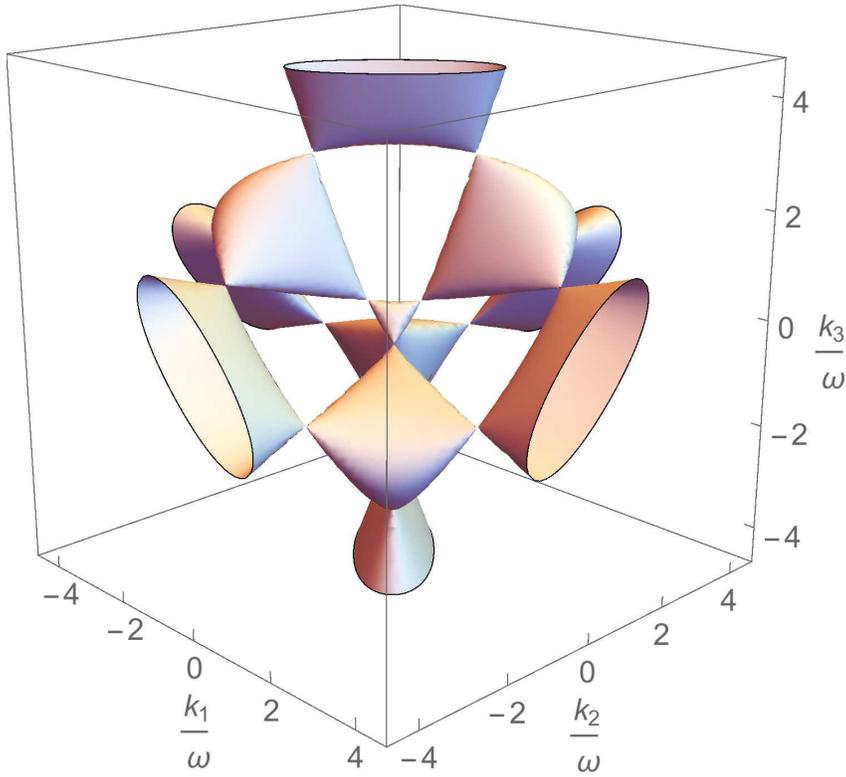}%{noinf.eps}
\caption{The permittivity $ \varepsilon^{ab}$, the impermeability $
  \mu^{-1}_{ab}$, and the magnetoelectric moduli $\gamma^{a}{}_{b}$ of
  the hypothetical metamaterial are specified in Eqs.(\ref{eq:noinf1},
  \ref{eq:noinf2}, \ref{eq:noinf3}), see Favaro et al.\
  \cite{Favaro:2015jxa}.}
\end{figure}

Schuller et al.\ \cite{Schuller} started {}from the doubly antisymmetric
and local and linear response tensor density $\chi^{ijkl}$, with
$^{(2)\!}\chi^{ijkl}=0$, which arises so naturally in
electrodynamics. They promoted it to a corresponding {\it area metric}
for an arbitrary $n$-dimensional manifold and extended their idea also
to string theory. For recovering a Fresnel type equation, they used a
generalized Tamm-Rubilar tensor, see \cite[Eq.(30)]{Schuller}. See
also some developments by Dahl \cite{Dahl:2011hb}. A fresh approach
has been set up recently by Ho \& Inami \cite{Ho:2015cza}.

%%%%%%%%%%%%%%%%%%%%%%%%%%%%%%%%%%%%%%%%%%%%%%%%%%%%%%%%%%%%%%%%%
\subsection{Eikonal method}
%%%%%%%%%%%%%%%%%%%%%%%%%%%%%%%%%%%%%%%%%%%%%%%%%%%%%%%%%%%%%%%%%

Itin \cite{Itin:2009aa} has developed a geometric optics analysis of
premetric Maxwell's equations on the basis of the eikonal ansatz
$F_{ij} = f_{ij} e^{i\sigma}$, where $f_{ij}$ are slowly changing
amplitude functions of the spacetime coordinates. By making use of
strictly algebraic methods, he found the following dispersion relation
(this is a synonym for the generalized Fresnel equation),
\begin{equation}\label{calc10}
  \epsilon_{ii_1\hspace{-0.5pt}i_2i_3}\epsilon_{jj_1j_2j_3}\Big(\chi^{i_1(ij)j_1}
  \chi^{i_2abj_2}+
  4\chi^{i_1(ia)j_1}\chi^{i_2(jb)j_2}\Big)\chi^{i_3cd\hspace{0.5pt}j_3}q_aq_bq_cq_d=0\,.
\end{equation}
Surprisingly, this expression looks different {}from the generalized
Fresnel equation (\ref{Fresnel}) cum (\ref{TammRubilar}). However,
Obukhov \cite[App.]{Itin:2009aa} was able to show, by means of heavy
algebraic manipulations, that (\ref{calc10}) and (\ref{Fresnel}),
apart {}from a numerical factor of $-72$, are just the same
expressions. This is, of course, what one had to expect.

Subsequently Itin has demonstrated that (\ref{calc10}) can be
transformed into different equivalent forms, the most symmetric form
probably being
\begin{equation}\label{calc11}
  \epsilon_{ii_1\hspace{-0.5pt}i_2i_3}\epsilon_{jj_1j_2j_3}\chi^{ii_1jj_1}
  \chi^{i_2abj_2}\chi^{i_3cd\hspace{0.5pt}j_3}q_aq_bq_cq_d=0\,.
\end{equation}
Exactly this dispersion relation, in a version using multiforms and
dyadics, was also found by Lindell \cite[Eq.(5.244)]{Lindell:2015},
bringing this subject of dispersion relations to a final close.

\subsection{Kummer tensor density of electrodynamics}

The problem of the generalized Fresnel equation/dispersion relation
for the local and linear medium was, as we saw, solved. Still, since
the year 1910 it was clear, see Bateman \cite{Bateman:1910}, that the
Fresnel {\it surface} should be a Kummer surface. 

Originally Kummer had started to investigate optical ray bundles of
second order, that is, those ray bundles that have, like in a
birefringent crystal, at one point two independent rays. The tool for
investigating these ray bundles was projective geometry---{\it a
  geometry without any metric}---in which a ray in 4-dimensional space
is projected into a real projective 3-dimensional space $RP_3$. These
investigations led Kummer around 1864 to a certain quartic wave
surface with, in general, 16 singular points \cite{Kummer:1864}, see
our Fig.\,6. This wave surface was later called Kummer surface by
Hudson \cite{Hudson_1903}, who wrote an authoritative review of such
surfaces. The proposal of the Kummer surface constituted a major
breakthrough in a field that is nowadays called algebraic geometry
(see \cite{Griffith:1978}). Some more details can be found in Baekler
et al.\ \cite{Baekler:2014kha}.

Favaro \cite{PhD}, in 2012, provided a critical and thorough
evaluation of the premetric electrodynamics with a local and linear
response law. Eventually, Baekler, Favaro, et al.\
\cite{Baekler:2014kha} introduced a new premetric tensor density of
4th rank, the Kummer tensor of local and linear electrodynamics, in
order to bring the property of Fresnel surfaces to be `Kummer' into
focus:
\begin{eqnarray}
\label{Kummerchi1}
{{{\cal K}^{ijkl}[\chi]}:=
{}^\diamond\chi^\diamond_{acbd}\,\chi^{aibj}\,\chi^{ckdl}\,.}
\end{eqnarray}
We immediately recognize the symmetry of this premetric definition,
\begin{equation}\label{pairsym}
{\cal K}^{ijkl}[\chi]={\cal K}^{klij}[\chi]\,.
\end{equation}
Note that the definition (\ref{Kummerchi1}) corresponds to Itin's most
symmetric case (\ref{calc11}) and to the analogous expression of
Lindell \cite{Lindell:2015}.

If ${\cal K}^{ijkl}[\chi]$ is understood as a symmetric $16\times 16$
matrix, it can be read off that it has 136 independent components. An
irreducible decomposition of the Kummer tensor density under the
$GL(4,R)$ was provided in \cite{Baekler:2014kha}. For the Tamm-Rubilar
tensor density, we find
\begin{eqnarray}\label{TR21}
   {\cal G}^{ijkl}[\chi] = \frac{1}{6}\,{\cal K}^{(ijkl)}[\chi]\,.
\end{eqnarray}
Accordingly, the totally symmetric piece of the Kummer tensor density
with its 35 independent components (it is one of the irreducible
pieces) can, up to a factor, observed in crystal optics. It has an
operational interpretation, exactly like ${\cal G}^{ijkl}$.  The name
``Kummer tensor'' was coined by Zund in the context of general
relativity \cite{Zund_1976}, see also Zund \cite{Zund_1969}. In his
definition, instead of $\chi^{ijkl}$, he used the Riemann curvature
tensor ${R}^{ijkl}=-{R}^{jikl}=-{R}^{ijlk}$ of general
relativity. However, ${R}^{ijkl}$, in contrast to $\chi^{ijkl}$, has
the additional symmetry ${R}^{ijkl}={R}^{klij}$ and, thus, carries
only 20 independent components (see the Post constraint!). Since
Kummer surfaces were first discovered in the context of {\it optics},
we find the new name Kummer tensor density perfectly appropriate in
electrodynamic considerations. So far, we were able to show that, for
vanishing skewon part, the Fresnel surfaces are Kummer surfaces,
exactly as Bateman \cite{Bateman:1910} had anticipated. This is a
decisive step forward in the understanding of Fresnel surfaces. Then,
our favorite form of the generalized Fresnel equation eventually turns
out to be 
\begin{eqnarray}\label{FresnelKummer}
  {\cal K}^{ijkl}[\chi]\,q_iq_jq_kq_l =0\,.
\end{eqnarray}
What is left to understand is the nature of the Fresnel surfaces for a
nonvanishing skewon part $^{(2)\!}\chi^{ijkl}\ne 0$ of the response
tensor density and the operational interpretation of the remaining
$136-35=101$ remaining components of the Kummer tensor density.

%\newpage
%%%%%%%%%%%%%%%%%%%%%%%%%%%%%%%%%%%%%%%%%%%%%%%%%%%%%%%%%%%%%%%%%%%%%
\section{Spacetime metric derived}\label{Metric}
%%%%%%%%%%%%%%%%%%%%%%%%%%%%%%%%%%%%%%%%%%%%%%%%%%%%%%%%%%%%%%%%%%%%%

Wave propagation in electrodynamics in materials with a local and
linear response law exhibits the phenomenon of birefringence (double
refraction), see Born \& Wolf \cite{Born:2001}. In the geometric
optics approximation, this is manifest in the fourth order Fresnel
equation (\ref{Fresnel}) for the wave covector, see our Figs.\,4 and
5. Such a picture is typical for anisotropic and magnetoelectric
materials.

Let us now consider the physical vacuum. It is natural to assume that
the physical vacuum is a non-birefringent continuum. In other words,
if we start with a local and linear medium and then, in case we want
to recover the physical vacuum, forbid birefringence, then we can hope
to recover the light cone of relativity theory---and thus the metric up
to a conformal factor. We will follow here partly our presentation in
\cite{Hehl:Brans}.

%%%%%%%%%%%%%%%%%%%%%%%%%%%%%%%%%%%%%%%%%%%%%%%%%%%%%%%%%%%%%%%%%%%
\subsection{Suppression of birefringence in vacuum: the light cone}
%%%%%%%%%%%%%%%%%%%%%%%%%%%%%%%%%%%%%%%%%%%%%%%%%%%%%%%%%%%%%%%%%%%

It was shown by L\"ammerzahl et al.\ \cite{Lammerzahl:2004ww,
  forerunner}, see also the review \cite{Hehl:2006}, that the
postulate of vanishing birefringence reduces the general Fresnel
equation to one for the light cone. Thus, up to a conformal factor,
the second rank spacetime metric $g_{ij}$ is recovered {}from the local
and linear fourth rank response tensor density $\chi^{ijkl}$ or
$\kappa_{ij}{}^{kl}$, respectively. These considerations have been
improved by Itin, Favaro, and Bergamin \cite{Itin:2005iv,G2,G3}.

Originally, we found the light cone by postulating certain reciprocity
constraints for the response tensor \cite{Birkbook}, see also Sec.\,9.4.
However, these arguments appear to be more mathematical, whereas
birefringence has a very intuitive meaning. Moreover, in the cosmos
birefringence is excluded with high accuracy, compare the observations
of Polarbear \cite{Ade:2015cao} and the discussion of Ni et al.\
\cite{Ni:2014qfa,Mei:2014iaa,Ni:2015poa}.

Looking at Fig.\,4, for instance, it is clear that we have to take
care that both shells in each Fresnel wave surface become identical
spheres. Then light propagates like in vacuum. For this purpose, we
can solve the quartic Fresnel equation (\ref{Fresnel}) with respect to
the frequency $q_0$, keeping the 3--covector $q_a$ fixed. One finds
four solutions, for the details please compare
\cite{Lammerzahl:2004ww,Itin:2005iv}. To suppress birefringence, one
has to demand {\it two conditions.} In turn, the quartic equation
splits into a product of two quadratic equations proportional to each
other. Thus, we find a light cone $g^{ij}(x)\,q_iq_j=0$ at each point
of spacetime.

Perhaps surprisingly, we derived also the {\it Lorentz signature,} see
\cite{Birkbook,Itin:2004qr,Itin:2008vk}.  Because of the importance of
this result, we will discuss it separately in Sec.\,12.

%%%%%%%%%%%%%%%%%%%%%%%%%%%%%%%%%%%%%%%%%%%%%%%%%%%%%%%%%%%%%%%%%%%%
\subsection{Dilaton, metric, axion}
%%%%%%%%%%%%%%%%%%%%%%%%%%%%%%%%%%%%%%%%%%%%%%%%%%%%%%%%%%%%%%%%%%%%

At the premetric level of our framework, we found that, besides the
principal piece, the skewon and the axion fields emerged. Only
subsequently the light cone was brought up. The skewon field was
phased out by our insistence of the vanishing birefringence in
vacuum. Accordingly, the {\it light cone} and the {\it axion field}
are the lone survivors of the collapse of birefringence.

By the light cone the metric $g_{ij}$ is only defined up to an
arbitrary function $\lambda(x)$, the dilaton field, see also Ni et
al.\ \cite{Ni:2014cca}. Thus, the vanishing birefringence leads to the
following response law:
\begin{equation}\label{antw}
  \check{\mathcal{H}}^{ij}=\boldsymbol{[}\underbrace{\lambda(x)}_{\text{dilaton}}
  \underbrace{\sqrt{-g}\,g^{ik}(x)\,g^{jl}(x)}_{\text{metric}}
  +\underbrace{\alpha(x)
  }_{\text{axion}} \epsilon^{ijkl}\,\boldsymbol{]} \,F_{kl}\,.
\end{equation}
In exterior calculus, it is even more compact by the use of the metric
dependent Hodge star $^\star$ operator: 
\begin{equation}\label{antw*}
  {H}=\boldsymbol{[}\lambda(x)\,^{\boldsymbol{\star}} +
  \alpha(x)\boldsymbol{]}F\,.
\end{equation}
We recognize that the three fields $\lambda(x)$, $g^{ij}(x)$, and
$\alpha(x)$ are all {\it descendants of electromagnetism} and they
emerge with a straightforward physical interpretation.

All experimental searches for the axion $A^0$ turned out to be
negative. Thus, at least for the time being, $\alpha=0$. Furthermore,
under conventional circumstances, the dilaton field seems to be
constant, $\lambda(x)= \lambda_{0}$, where $\lambda_{0}$ is the vacuum
admittance with a value of about $ 1/377\hspace{1pt}\Omega$. Accordingly, we
end up with the Maxwell-Lorentz law
\begin{equation}
  \mathcal{H}^{ij}%=\lambda_0\sqrt{-g}\,g^{ik}(x)\,g^{jl}(x)\,F_{kl}
  =  \lambda_0\sqrt{-g}\,F^{ij}\qquad \hbox{or}\qquad  H=\lambda_0\,^\star F\,,
\end{equation}
which coincides with  (\ref{const}).

%%%%%%%%%%%%%%%%%%%%%%%%%%%%%%%%%%%%%%%%%%%%%%%%%%%%%%%%%%%%%%%%%%%%%
\section{The sign of the electromagnetic energy, the Lenz
rule, and  the emergence of the Lorentz signature}\label{Lenz}
%%%%%%%%%%%%%%%%%%%%%%%%%%%%%%%%%%%%%%%%%%%%%%%%%%%%%%%%%%%%%%%%%%%%%

The premetric Maxwell's equations, $dH = J$ and $dF = 0$, are universally
valid in a four-dimensional manifold, on which time and space coordinates
are arbitrarily introduced by means of the $(1 + 3)$-foliation \cite{Birkbook}.
Any physical object, say a form $\Psi$ of a rank $p$, can be uniquely
decomposed $\Psi = d\tau\wedge\Psi_\bot + \underline{\Psi}$ into the
longitudinal $(p -1)$-form $\Psi_\bot$ and the transversal $p$-form
$\underline{\Psi}$.

Although the $(1+3)$ splitting is unique, the interpretation of the
longitudinal and transversal parts of physical object in Maxwell's
theory admits a certain freedom. The resulting three-dimensional forms
of premetric electrodynamics correspond to the different basic
physical laws such as positivity versus negativity of the energy,
attraction versus repulsion between two charges, and so on.  In this
section we present a brief analysis of these different possibilities
and give a justification of the sign conventions presented in
Sec.\,\ref{Maxwell}. For the proofs and the details, see
\cite{Birkbook,Itin:2004qr,Itin:2008vk}.

The analysis of the $(1+3)$-decomposition of the electric current $J$ shows
that all the possibilities of the sign factors are merely conventional and
the splitting has the ordinary form
 \begin{equation}\label{decomp8}
   J=- j\wedge d\tau+\rho\,,
\end{equation}
where the standard notations for the 3-dimensional current $j:=J_\bot$
and charge density $\rho:=\underline{J}$ are used. The uniqueness here is
due to the fact that for a 3-form  $J$, the spatial exterior differential
of the transversal 3-form $\rho$ vanishes identically.

Turning to the excitation 2-form $H$, we arrive at the decomposition
\begin{equation}\label{Hdecomp}
  H=h_{\tt T}\,{\mathcal H}\wedge d\tau +\,{\mathcal   D}\,.
\end{equation}
The sign factor $ h_{\tt T} = \pm 1$ remains undetermined, reflecting the
freedom to identify the magnetic excitation ${\mathcal H}$ either with
the longitudinal part $H_\bot$ or with $-H_\bot$. Thus, the inhomogeneous
Maxwell equations read
\begin{equation}\label{inhomMax}
    %\left\{ \begin{array}{l}
  h_{\tt T}\,\underline{d}\,{\mathcal H}+\,\dot{\mathcal D}=-j\,,\qquad
  \underline{d}\,\mathcal D=\rho \,.%\end{array} \right.
\end{equation}

Quite similarly, the decomposition of the 2-form $F$ of the
electromagnetic field strength is given by
    \begin{equation}\label{Fdecomp}
      F=f_{\tt T}\,E\wedge d\tau+B\,.
\end{equation}
and the field equations by
\begin{equation}\label{homMax2}
  f_{\tt T}\,\underline{d}\,E+\dot{B} = 0\,,\qquad
  \underline{d}\,B=0\,,
\end{equation}
with another undetermined sign factor $f_{\tt T} = \pm 1$.

In order to fix the free factors $ h_{\tt T}$ and $f_{\tt T}$ we turn
to an additional ingredient of premetric electrodynamics---the
energy-momentum current 3-form, see \cite{Birkbook}. The ``time''
(transversal) component of this current 3-form represents the energy
density of the field. In our setting it takes the form
\begin{equation}\label{em-energy}
  u=\frac 12 f_{\tt T}\, E\wedge {\mathcal D} -\frac 12 h_{\tt T}\,
  {\mathcal H}\wedge B\,.
\end{equation}
Up to this point, {\it everything was premetric.} Now we have to take
recourse to the vacuum constitutive relations. Substituting the
Maxwell-Lorentz constitutive relation {\it for vacuum} (\ref{constM})
into (\ref{em-energy}), we obtain
\begin{equation}\label{em-energy*}
  u=\frac 12\left\{ f_{\tt T}\varepsilon_0\sqrt{g}\left(  g^{ab}E_aE_b\right)\,
    -\frac {h_{\tt T}}{\mu_0}\frac 1{\sqrt{g}}\,\left(
      g_{ab}B^aB^b\right)\right\}\epsilon\,.
\end{equation}
The first term in (\ref{em-energy*}) represents the pure electric part
of the energy density while the second part is the pure magnetic part.
We requite that both parts of the energy are non-negative. With the
Euclidean 3d metric $g_{ab}$, we arrive to the conditions
\begin{equation}\label{coef-con}
f_{\tt T}\varepsilon_0>0\,,\qquad {h_{\tt T}}{\mu_0}<0\,.
\end{equation}
We then readily recognize that for positive values of the electric and
magnetic constants $\varepsilon_0$ and $\mu_0$, the ordinary sign
factors $f_{\tt T} = +1$ and $h_{\tt T} = -1$ of Maxwell's
electrodynamics are recovered. Nevertheless, one can speculate about
various possibilities arising for negative values of the vacuum
constants.

Let us list some basic physical consequences of the inequalities
(\ref{coef-con}):

\vspace{0.2cm}

\noindent$\bullet$ {\bf Dufay's law:} In our setting, the electric
force between two static charges comes with an additional factor of
$f_{\tt T}$, see \cite{Itin:2004qr}. In particular, $f_{\tt T}=+1$
yields attraction between opposite charges and repulsion between
charges of the same sign: this is Dufay's law. In contrast, $f_{\tt
  T}=-1$ yields {\it anti-}Dufay law with attraction between charges
of the same sign and repulsion between opposite charges.

\vspace{0.2cm}

\noindent$\bullet$ {\bf Lenz's rule:} For $h_{\tt T}=-1$, we have the
ordinary relative sign in Faraday's law, that is usually described by
Lenz's rule. It is responsible for the pulling of a ferromagnetic core
into a solenoid independently on the direction of the current. For
$h_{\tt T}=+1$, the situation is opposite and the corresponding {\it
  anti-}Lenz rule holds.

\vspace{0.2cm}

\noindent$\bullet$ {\bf Lorentz signature:} By considering the
electromagnetic waves in our construction we arrive to the ordinary
wave equation for the field $E$, for instance. The sign factors in
this equation are completely canceled and the square of the wave
velocity is given by the usual formula
$c^2=1/(\varepsilon_0\mu_0)$. The form of the wave equation manifests
the properties of the underlying geometry.  Thus, the ordinary case
$\varepsilon_0\mu_0>0$ corresponds to the Lorentzian signature of the
$4$-metric, whereas the case $\varepsilon\mu<0$ corresponds to the
Euclidean signature.\medskip

Our analysis shows that in the framework of Kottler's premetric
approach we are able to derive the basis physical properties of
electromagnetism in vacuum. For this, the ordinary vacuum constitutive
relation must be assumed and the positivity of the electric and
magnetic energy must be required.

Let us extend now the constitutive relation Eq.(\ref{constM}) to
isotropic media. As a result, the vacuum constants are replaced by the
permittivity $\varepsilon$ and permeability $\mu$ parameters of the
medium. On can even speculate that the media parameters are not
necessary positive, which can be the case for the artificially
designed materials -- metamaterials. We collect all possible sign
options with their physical consequences in the following
table:\bigskip

{\bf Table IV.} Classification of physical media by sign factors%Sign
                                %rules for ordinary and hypothetical media
%\begin{center}
\begin{table}[h]
%\begin{center}
\begin{tabular}{|l|l|l|l|}
  \hline
$f_{\tt T}=+1\,,h_{\tt T}=-1 $&Lenz rule&Dufay law&Lorentzian\\
$\varepsilon>0\,,\,\quad \mu>0$&&&signature\\
 \hline
$f_{\tt T}=-1\,,h_{\tt T}=-1 $&Lenz rule&anti-Dufay law&Euclidean\\
$\varepsilon<0\,,\,\quad \mu>0$&&&signature\\
 \hline
$f_{\tt T}=+1\,,h_{\tt T}=+1 $&anti-Lenz rule&Dufay law&Euclidean\\
$\varepsilon>0\,,\,\quad \mu<0$&&&signature\\
 \hline
$f_{\tt T}=-1\,,h_{\tt T}=+1 $&anti-Lenz rule&anti-Dufay law&Lorentz\\
$\varepsilon<0\,,\,\quad \mu<0$&&&signature\\
 \hline
% \end{center}
  \end{tabular}
\end{table}
%\end{center}

This analysis is based on the assumption of the positivity of the
electric and magnetic energy densities. In contrast, the ordinary
description of metamaterials allows for the negative energies. In this
case, the standard Dufay and Lenz rules are recovered in our model,
too.

Let us examine now the most general premetric linear constitutive
relation (\ref{const1})--(\ref{const2}).  We find
\begin{equation}\label{ED}
 E\wedge {\mathcal D}=E_aD^a\,\epsilon=\left(-\,{\mathfrak A}^{ba}\,E_aE_b
 + {\mathfrak D}_{b}{}^a\,E_aB^b\right)\epsilon\,,
\end{equation}
and
\begin{equation}\label{HB'}
  {\mathcal H}\wedge B=H_aB^a\,\epsilon=\left(-\,{\mathfrak C}^{b}{}_a
    \,E_b B^a+ {\mathfrak B}_{ba}\,B^bB^a\right)\epsilon\,
\end{equation}
Consequently, the energy density (\ref{em-energy*}) takes the form
\begin{equation}\label{em-energy1}
  u=-\frac 12\left( f_{\tt T}\, {\mathfrak A}^{ab}\,E_aE_b + h_{\tt
      T}\, {\mathfrak B}_{ab}\,B^aB^b-\left(f_{\tt T}\,
      {\mathfrak D}_{b}{}^a+h_{\tt T}\,{\mathfrak
        C}^{a}{}_b\right)E_aB^b\right)  \epsilon\,.
\end{equation}
In the first two terms we recognize the electric and magnetic energies,
respectively. For the standard Lorentzian signature with $ f_{\tt
  T}=+1$ and $h_{\tt T}=-1$, the matrix ${\mathfrak A}^{ab}$ must be
negative definite and the matrix ${\mathfrak B}_{ab}$ must be positive
definite.  This result yields positive definiteness of the matrices
$\varepsilon_{ab}$ and $\mu_{ab}$.

%%%%%%%%%%%%%%%%%%%%%%%%%%%%%%%%%%%%%%%%%%%%%%%%%%%%%%%%%%%%%%%%%%%%%
\section{Concluding remarks}\label{Concl}
%%%%%%%%%%%%%%%%%%%%%%%%%%%%%%%%%%%%%%%%%%%%%%%%%%%%%%%%%%%%%%%%%%%%%

We should mention that by introducing generalized concepts of stress
and hyperstress, one can set up a kind of a {\it premetric elasticity}
theory, as shown by Gronwald et al.\ \cite{Elba}. Only if one relates
this framework to the classical Euler stress, which is symmetric, one
has to eventually to introduce the metric tensor. Moreover, Favaro
(private communication) has pointed out that, in non equilibrium
statistical mechanics, the {\it Fokker-Planck equation} can be
formulated in a premetric way.

The Kottler program was intended to remove the metric {}from the
fundamental equations in physics as far as possible.  Since the metric
is the gravitational potential in general relativity theory, Kottler's
program does not apply {here. However, in the teleparallel
  version of gravity, it seems possible to carry through the Kottler
  program, although the details will have to be seen. In classical
  electrodynamics, Kottler's path led to a resounding success and,}
see Sec.\,11, to a new understanding of the emergence of the metric
tensor together with a dilaton, an axion, and a skewon field. All
these fields, together with the metric, come into existence in an
electrodynamic context. These remarkable facts are not yet completely
understood. In particular, only a first attempt was started to develop
a theory for a skewon field \cite{Hehl:2005xu}, for which some cosmic
limits were set by Ni \cite{Ni:2013uwa}.

%\appendix

%%%%%%%%%%%%%%%%%%%%%%%%%%%%%%%%%%%%%%%%%%%%%%%%%%%%%%%%%%%%%%%%%%%%%
\subsection*{Acknowledgments}
%%%%%%%%%%%%%%%%%%%%%%%%%%%%%%%%%%%%%%%%%%%%%%%%%%%%%%%%%%%%%%%%%%%%%

%We would like to thank Wei-Tou Ni (Hsin-chu) for the invitation to
%contribute an article to his electrodynamics issue of IJMPD. 
We would like to extend our sincere thanks to Wei-Tou Ni (Hsin-chu)
for his encouragement to write up this article.  Quite a number of
colleagues contributed appreciably to our understanding of the subject
of our investigation. We are most grateful to the following people: To
Alberto Favaro (London) for many detailed discussions on the
foundations of electrodynamics and on Kummer surfaces; to Enzo Tonti
(Trieste) for advice on the operational interpretation of
electrodynamics; to Naresh Dadhich (Pune) and to Claus Kiefer
(Cologne) for insight on fundamental constants; to Reinhard Meinel
(Jena) on gravitomagnetism; to Hubert Goenner (G\"ottingen) and,
particularly, to Tilman Sauer (Mainz) (G\"ottingen) for help in
historical questions; to Daniel Braun (T\"ubingen) for discussions on
the precision with which the speed of light can be measured; to Ari
Sihvola (Espoo) for explanations to \cite{Serdyukov:2001}; and to
Achim Rosch (Cologne) for decisive hints to copper selenide.

%\begin{footnotesize}

%\end{footnotesize}
\end{document}